\documentclass[11pt]{JHEP}
\expandafter\ifx\csname amssym.def\endcsname\relax \else\endinput\fi
\expandafter\edef\csname amssym.def\endcsname{%
       \catcode`\noexpand\@=\the\catcode`\@\space}
\catcode`\@=11
\def\undefine#1{\let#1\undefined}
\def\newsymbol#1#2#3#4#5{\let\next@\relax
 \ifnum#2=\@ne\let\next@\msafam@\else
 \ifnum#2=\tw@\let\next@\msbfam@\fi\fi
 \mathchardef#1="#3\next@#4#5}
\def\mathhexbox@#1#2#3{\relax
 \ifmmode\mathpalette{}{\m@th\mathchar"#1#2#3}%
 \else\leavevmode\hbox{$\m@th\mathchar"#1#2#3$}\fi}
\def\hexnumber@#1{\ifcase#1 0\or 1\or 2\or 3\or 4\or 5\or 6\or 7\or 8\or
 9\or A\or B\or C\or D\or E\or F\fi}

\font\tenmsa=msam10
\font\sevenmsa=msam7
\font\fivemsa=msam5
\newfam\msafam
\textfont\msafam=\tenmsa
\scriptfont\msafam=\sevenmsa
\scriptscriptfont\msafam=\fivemsa
\edef\msafam@{\hexnumber@\msafam}
\mathchardef\dabar@"0\msafam@39
\def\dashrightarrow{\mathrel{\dabar@\dabar@\mathchar"0\msafam@4B}}
\def\dashleftarrow{\mathrel{\mathchar"0\msafam@4C\dabar@\dabar@}}

\def\ulcorner{\delimiter"4\msafam@70\msafam@70 }
\def\urcorner{\delimiter"5\msafam@71\msafam@71 }
\def\llcorner{\delimiter"4\msafam@78\msafam@78 }
\def\lrcorner{\delimiter"5\msafam@79\msafam@79 }
\def\yen{{\mathhexbox@\msafam@55}}
\def\checkmark{{\mathhexbox@\msafam@58}}
\def\circledR{{\mathhexbox@\msafam@72}}
\def\maltese{{\mathhexbox@\msafam@7A}}
\def\circledS{{\mathhexbox@\msafam@73}}
\newsymbol\square 1003

\font\tenmsb=msbm10
\font\sevenmsb=msbm7
\font\fivemsb=msbm5
\newfam\msbfam
\textfont\msbfam=\tenmsb
\scriptfont\msbfam=\sevenmsb
\scriptscriptfont\msbfam=\fivemsb
\edef\msbfam@{\hexnumber@\msbfam}
\def\Bbb#1{{\fam\msbfam\relax#1}}
\def\widehat#1{\setbox\z@\hbox{$\m@th#1$}%
 \ifdim\wd\z@>\tw@ em\mathaccent"0\msbfam@5B{#1}%
 \else\mathaccent"0362{#1}\fi}
\def\widetilde#1{\setbox\z@\hbox{$\m@th#1$}%
 \ifdim\wd\z@>\tw@ em\mathaccent"0\msbfam@5D{#1}%
 \else\mathaccent"0365{#1}\fi}
\font\teneufm=eufm10
\font\seveneufm=eufm7
\font\fiveeufm=eufm5
\newfam\eufmfam
\textfont\eufmfam=\teneufm
\scriptfont\eufmfam=\seveneufm
\scriptscriptfont\eufmfam=\fiveeufm

\csname amssym.def\endcsname

\renewcommand{\nc}{\newcommand}
\newcommand{\rnc}{\renewcommand}

\nc{\be}{\begin{equation}}
\nc{\ee}{\end{equation}}
\nc{\bea}{\begin{eqnarray}}
\nc{\eea}{\end{eqnarray}}

\nc{\trac}[2]{{\textstyle\frac{#1}{#2}}}

\nc{\ex}[1]{\mbox{e}^{\,\textstyle#1}}

\nc{\CC}{\Bbb{C}}
\nc{\HH}{\Bbb{H}}
\nc{\PP}{\Bbb{P}}
\nc{\RR}{\Bbb{R}}
\nc{\ZZ}{\Bbb{Z}}
\nc{\II}{\Bbb{I}}
\nc{\EE}{\Bbb{E}}

\rnc{\a}{\alpha}
\nc{\ab}{\alpha^{*}}
\nc{\al}{\a^{l}}
\rnc{\d}{\delta}
\nc{\ga}{\gamma}
\nc{\la}{\lambda}
\nc{\lal}{\la_{l}}
\nc{\f}{\phi}
\nc{\fb}{\bar{\phi}}
\nc{\p}{\psi}

\nc{\e}{\varepsilon}
\nc{\eb}{\bar{\e}}
\rnc{\c}{\chi}
\nc{\cb}{\bar{\chi}}
\nc{\eps}{\epsilon}
\rnc{\t}{\theta}
\nc{\tb}{\bar{\theta}}
\nc{\om}{\omega}

\rnc{\P}{\Psi}
\nc{\pl}{\P_{L}}
\nc{\pdr}{\P^{\dag}_{R}}
\nc{\G}{\Gamma}
\nc{\Ga}{\Gamma}

\nc{\sig}{\sigma}
\nc{\sk}{\sigma_{k}}
\nc{\sa}{\sigma_{a}}
\nc{\Bb}{\bar{B}}

\nc{\symx}{\circledS}

\nc{\Q}{\bar{Q}}
\nc{\C}{{\cal A}/{\cal G}}                
\nc{\A}[1]{{\cal A}^{#1}/{\cal G}^{#1}}  
\nc{\RC}{{\cal R}_{\C}}                 
\nc{\RM}{{\cal R}_{\M}}                
\nc{\RX}{{\cal R}_{X}}
\nc{\RY}{{\cal R}_{Y}}

\nc{\ad}{\mathop{\mbox{ad}}\nolimits}
\nc{\tr}{\mathop{\mbox{tr}}\nolimits}
\nc{\Tr}{\mathop{\mbox{Tr}}\nolimits}
\nc{\Det}{\mathop{\mbox{Det}}\nolimits}
\rnc{\det}{\mathop{\mbox{det}}\nolimits}
\nc{\rk}{\mathop{\mbox{rk}}\nolimits}
\nc{\diag}{\mbox{diag}}
\nc{\ra}{\rightarrow}
\nc{\Ra}{\Rightarrow}
\nc{\LRa}{\Leftrightarrow}
\nc{\lra}{\leftrightarrow}
\nc{\ot}{\otimes}
\rnc{\ss}{\subset}
\nc{\ul}{\underline}
\nc{\nul}{\noindent\underline}
\nc{\non}{\nonumber\\}
\rnc{\S}{\Sigma}
\nc{\tp}{2\pi i}
\nc{\del}{\partial}
\nc{\dbar}{\bar{\del}}
\nc{\dx}{\dot{x}}
\nc{\zb}{\bar{z}}

\rnc{\lg}{\log g^{2}}
\nc{\lv}{\log V_{s}}
\nc{\vs}{V_{s}}
\rnc{\ln}{\log \N}
\nc{\ls}{\ell_{s}}
\nc{\N}{{\cal N}}
\nc{\M}{{\cal M}}
\nc{\F}{{\cal F}}
\nc{\E}{{\cal E}}
\rnc{\P}{{\cal P}}
\nc{\I}{{\cal I}}
\nc{\IIt}{$\widetilde{\mbox{II}}$}
\nc{\gst}{\widetilde{g_{s}}}
\nc{\gsh}{\widehat{g_{s}}}
\nc{\lsh}{\widehat{\ls}}
\nc{\rllh}{\widehat{R_{11}}}
\nc{\lph}{\widehat{\ell_{P}}}
\nc{\mnd}{M_{Nd}}

\nc{\nam}{\nabla_{\mu}}
\nc{\nan}{\nabla_{\nu}}

\nc{\mat}[4]{\left(\begin{array}{cc}#1&#2\\#3&#4\end{array}\right)}
 
\nc{\rb}[1]{\overline{\mathbf{#1}}}

\nc{\gi}{\gamma_{i}}
\nc{\gj}{\gamma_{j}}

\nc{\chap}[1]{{\clearpage}%
\begin{center}%
{\noindent\underline{\large\sc #1}}{\addcontentsline{toc}{section}{#1}}%
\end{center}%
{\vspace*{0.3cm}}}

\newcommand{\ba}{\begin{eqnarray}}
\newcommand{\ea}{\end{eqnarray}}

\title{Killing Spinors and SYM on Curved Spaces}
\author{Matthias Blau\\Abdus Salam ICTP\\ Strada Costiera 11\\ 
34014 Trieste, Italy\\ \email{mblau@ictp.trieste.it}}

\abstract{
We construct two families of globally supersymmetric
counterparts of standard Poincar\'e supersymmetric SYM theories on curved
space-times admitting Killing spinors, in all dimensions less than six
and eight respectively. The former differs from the standard theory only
by mass terms for the fermions and scalars and modified supersymmetry
transformation rules, the latter in addition has cubic Chern-Simons like
couplings for the scalar fields. We partially calculate the supersymmetry
algebra of these models, finding R-symmetry extensions proportional
to the curvature. We also show that generically these theories have no
continuous Coulomb branch of maximally supersymmetric vacua, but that
there exists a half-BPS Coulomb branch approaching the standard Coulomb
branch in the Ricciflat limit.}

\preprint{{\tt hep-th/0005098}}

\keywords{Brane Dynamics in Gauge Theories, Extended Supersymmetry}

\begin{document}
\baselineskip 4.0ex

\section{Introduction}

It is well known that Poincar\'e supersymmetric gauge theories retain
a certain fraction of their supersymmetry when placed on Ricci flat
manifolds $M$ admitting covariantly constant spinors, simply by using
these parallel spinors as the supersymmetry parameters. For the same
reason string theory compactifications on such manifolds lead to 
space-time supersymmetry.

{}From the string or supergravity theory point of view it is almost equally
natural to consider (maximally) supersymmetric compactifications of the
form $M_{1}\times M_{2}$ where this time the $M_{i}$ are required to
be Einstein manifolds admitting Killing spinors rather than covariantly
constant spinors. 

It is therefore natural to ask if super-Yang-Mills (SYM) theories retain
some global supersymmetry when placed on backgrounds admitting Killing
spinors. For instances, this question arises in the context of the
AdS/CFT correspondence \cite{jm,gkp,ew,adsrev} when considering curved
wrapped D-branes, as e.g.\ in \cite{ew2,yi,grst}. It also ought to arise,
for the same reason as in the case of branes wrapped over supersymmetric
cycles of manifods admitting parallel spinors (see e.g.\ \cite{bsv,bt1}),
in the context of AdS-calibrations studied in \cite{jggp,gpt}.

Morally speaking, by virtue of the existence of Killing spinors,
globally supersymmetric SYM theories should exist on such manifolds, and
it should be possible to deduce their existence and properties directly,
i.e.\ without having to pass through supergravity and the possibly arduous
task of studying fluctuations around a given (perhaps not even maximally)
supersymmetric background. 

It appears to be almost folklore knowledge that for the four-dimensional
SYM theories addition of a suitable mass term for the scalars in the
vector multiplet is sufficient to ensure supersymmetry on a background
with Killing spinors. However, I am not aware of any general and systematic,
i.e.\ not tied to a particular dimension, discussion of these 
matters.

Here, in addition to reproducing these results for $n=4$, we will find
two families of Killing SYM theories for $n\neq 4$, both of them with
the same field content as their Poincar\'e supersymmetric counterparts
but with different actions and (generically) different supersymmetry
transformation laws. From the results one can see in retrospect that the
four-dimensional case (with equal masses for all the scalars, no other
scalar potential terms, no mass term for the fermions) is sufficiently
special to preclude a straightforward extrapolation to other dimensions.

One of these families of theories, given in (\ref{l1}),
has the presumably unsurprising property
of differing from the flat space theory by mass terms for the scalar
fields and (unless the space-time dimension is $n=4$) fermions. I would
suspect that these theories can be readily extracted from the supergravity
literature.  However, even one of the simplest members of this family of
theories we will find, namely the $N=2$ theory on $AdS_{5}$, was only
constructed very recently in \cite{sh}, so perhaps these theories are
not so well known after all.

The other family, given in (\ref{l3}), existing in all dimensions $n\leq
7$, has the more curious feature of requiring Chern-Simons-like cubic
couplings of the scalar fields for supersymmetry and appears to be new.

One unexpected
consequence of this is the existence of two inequivalent supersymmetric
curved space counterparts of the three-dimensional $N=4$ SYM theory on
locally AdS spaces: one with with fermionic and bosonic mass terms and
modified supersymmetry transformation rules, the other with the same
supersymmetry transformation rules as in flat space but with a cubic
interaction term for the scalars instead of a mass term.

If these Killing SYM theories are realized as world volume theories
of certain curved D-branes - and the role of wrapped branes e.g.\ in
studies of the AdS/CFT correspondence certainly suggests that they
should be thought of as being equipped with a supersymmetric world
volume dynamics - then certainly the fundamental properties of these
theories, supersymmetric vacua, BPS configurations etc., need to be
understood. Here we will just discuss one simple but intriguing aspect
of these theories, namely the counterpart of what is usually called the
Coulomb branch. What we will find is that the structure of the vacua
with unbroken supersymmetries in these theories differs quite markedly
from that in the Poincar\'e supersymmetric theories - e.g.\ 
in the sense that
generically there is no continuous family of maximally 
supersymmetric vacua, i.e.\ all the flat directions of the potential
are lifted by a contribution to the potential induced by the curvature.

This in itself may not be terribly surprising, given the known results
about other quantum field theories in AdS space-times.  However, it
certainly calls for a reappreciation of these issues in the context of
brane dynamics.

As signs are crucial when it comes to checking supersymmetry, section 2
and an appendix serve to establish the conventions and notation and to
provide some background information regarding supersymmetry variations in
curved backgrounds and Killing spinors. In section 3, the two classes of
theories mentioned above are described, and in section 4 the supersymmetry
algebra in these models is (partially) calculated. Section 5 contains
some sample calculations in these models, dealing mainly with the absence
of a maximally supersymmetric Coulomb branch and the existence of a half-BPS
Coulomb branch. 

There are a large number of open issues, e.g.\ a more conceptual
understanding of the existence of these theories (which here have been
constructed more or less by brute force), and their superalgebraic
underpinning, the study of the corresponding quantum theories, spaces
of vacua, BPS configurations, application to worldvolume theories of
curved D-branes, etc. Work on these and related issues (the original
motivation for looking at (and hence first for) these theories was
part of an attempt to find a topological counterpart of the AdS/CFT
correspondence) is in progress, and I will briefly come back to these
issues in the concluding section 6.


\acknowledgments

I am grateful to Jose Figueroa-O'Farrill, Edi Gava, K.S.\ Narain, Martin
O'Loughlin, Seif Randjbar-Daemi and George Thompson for discussions
and suggestions at various stages of this work and for encouraging me
to finally write up these results.  This work was supported in part by
the EC under the TMR contract ERBFMRX-CT96-0090.

\section{Background}

\subsection{SYM Theories in Flat Space}

We will consider the $N=1$ SYM theories in $d=2+1, 3+1, 5+1$ and $9+1$
dimensions as well as their dimensional reductions to $n\leq d$ dimensions.
This dimensional reduction could be along space-like directions to
produce the standard Minkowski signature SYM theories, but it could
also involve the time-direction to give rise to hermitian SYM actions
in Euclidean signature \cite{bt2,qmc2,bvn}. Thus in particular these theories
include the $N=2$ and $N=4$ theories in $n=3+1$ as well as their Euclidean
counterparts.

Quite generally, for all these theories the Lagrangian in $d$ 
or $n$ dimensional flat space can be written in the compact form 
\be
L_{SYM}=-\trac{1}{2}F_{MN}F^{MN} + \bar{\Psi}\Ga^{M} D_{M}\Psi\;\;.
\label{lsym}
\ee
Here the following conventions have been used:  
\begin{itemize}
\item Capital indices $L,M,N,\ldots$ run from $0$ to $d-1$.
\item The gauge fields $A_{M}$ and $\Psi$ only depend on the 
      coordinates $x^{\mu}$, $\mu=0,\ldots,n-1 $ or $\mu=1,\ldots,n$
      depending on whether one performs a space or time reduction.
      Thus $A_{\mu}$ is an $n$-dimensional 
      gauge field and the remaining $(d-n)$ components
      $A_{m}\equiv \phi_{m}$ are scalar fields transforming as a vector under
      the manifest R-symmetry group $SO(d-n)$ or $SO(d-n-1,1)$.
\item A trace is implicit in (\ref{lsym}) for the interacting
      (non-Abelian) theories, the fields transforming in the 
      adjoint representation of the gauge group $G$, 
\be A_{M} = A_{M}^{i}T_{i}\;\;,\;\;\;\;\;\;\Psi = \Psi^{i}T_{i}\;\;.
\ee
      These Lie algebra indices will usually be suppressed in the following.
\item $A_{M}$ will be taken to be anti-hermitian, so that the field strength
      tensor is
\be
F_{MN}=\del_{M}A_{N}-\del_{N}A_{M} + [A_{M},A_{N}]
\ee
      (no factors of $i$).
\item The $\Ga^{M}$ are $d$-dimensional unitary gamma matrices and satisfy
\be
\{\Ga_{M},\Ga_{N}\} = \eta_{MN}
\ee
with
\be
\eta_{MN} = \diag(-1,\underbrace{+1,\ldots,+1}_{d-1})\;\;.
\ee
\item $\Psi$ is an anticommuting
spinor in $d$ dimensions satisfying the
condition 
\be
\begin{array}{cl}
d=2+1: & \mbox{Majorana} \\ 
d=3+1: & \mbox{Majorana or Weyl} \\ 
d=5+1: & \mbox{Weyl} \\ 
d=9+1: & \mbox{Majorana-Weyl}
\end{array}
\label{psi}
\ee
\item $\bar{\Psi}$ is the Dirac adjoint of $\Psi$ defined by 
\be
\bar{\Psi} = \Psi^{\dagger} A_{-}\;\;,
\ee
where $A_{-}=\Gamma_{0}$ satisfies
\be
\Gamma_{M}^{\dagger} = - A_{-}\Gamma_{M} A_{-}^{-1}\;\;.
\ee
\item $D_{M}$ is the gauge covariant derivative,
\bea
D_{\mu}\Psi &=& \del_{\mu}\Psi + [A_{\mu},\Psi]\non
D_{m} \Psi &=& [\phi_{m},\Psi]\;\;.
\eea
\end{itemize}
With these conventions, and the rule
\be
(\chi^{\dagger}\psi)^{\dagger} = - \psi^{\dagger}\chi
\ee
for anticommuting spinors $\chi, \psi$, 
the above action is hermitian. Explicitly it reads
\bea
L_{SYM} &=& 
-\trac{1}{2}F_{\mu\nu}F^{\mu\nu} -
D_{\mu}\phi_{m} D^{\mu}\phi^{m} 
- \trac{1}{2}[\phi_{m},\phi_{n}] [\phi^{m},\phi^{n}]\non
&+& \bar{\Psi}\Ga^{\mu} D_{\mu}\Psi
+ \bar{\Psi}\Ga^{m} [\phi_{m},\Psi] \;\;.
\label{lsym2}
\eea
In flat space it
is invariant under the supersymmetry transformations
\bea
\delta A_{M}^{i} &=& (\bar{\e}\Ga_{M}\Psi^{i} 
- \bar{\Psi}^{i}\Ga_{M}\e)\non
\delta \Psi^{i}  &=&  \Ga^{MN}F^{i}_{MN}\e \non
\delta \bar{\Psi}^{i}  &=& - \bar{\e}\Ga^{MN}F^{i}_{MN} 
\label{delta}
\eea
(modulo total derivatives) when $\e$ is a constant
spinor also satisfying the condition (\ref{psi}). Here 
\be
\Ga^{MN} = \trac{1}{2}[\Ga^{M},\Ga^{N}]\;\;. 
\ee
In the non-Abelian case, 
vanishing of the quartic fermionic terms arising from the variation of
the gauge field in the fermion kinetic term requires a Fierz identity
to hold, which is satisfied by virtue of the conditions (\ref{psi}). The
free theories are invariant under (\ref{delta}) without this requirement.

For brevity we will frequently refer to the dimensional reduction of the
$d$-dimensional $N=1$ theory to $n$ dimensions as the $(d,n)$ theory.
Thus the $(10,4)$ theory is $N=4$ SYM in four dimensions and e.g.\ 
$(6,5)$ refers to the five dimensional $N=2$ theory with one Dirac
spinor (actually two symplectic Majorana spinors, hence $N=2$) 
and one real scalar in addition to the five-dimensional gauge
field. We will mostly consider standard space-like reductions, but
following the procedure outlined in \cite{bt2} one can also obtain
Euclidean SYM theories by performing the dimensional reduction along
the time-direction. These will be discussed seperately in section 3.4.

\subsection{Supersymmetry Variations in Curved Space}

Let us now consider what happens when one tries to place these
theories (after the appropriate dimensional reduction) on a curved
background. To be specific, denote by $(M,g)$ a (pseudo-)Riemannian
$n$-dimensional spin manifold with metric $g_{\mu\nu}$. 

There is of course 
no problem with writing down the action (\ref{lsym}) on $M$ by
introducing a vielbein $e^{a}_{\;\mu}$, a spin connection 
$\omega^{ab}_{\mu}$, etc. Just to further pin down the conventions,
the spin connection part of the covariant derivative is
\be
\nabla_{\mu}\Psi = \del_{\mu}\Psi + \trac{1}{4}\Gamma_{ab}\omega^{ab}_{\mu}\Psi 
\;\;.\label{nabla}
\ee
The real issue is whether this theory will have any supersymmetry, the
point being that constant spinors $\e$ will in general not
exist on $M$ while using non-constant supersymmetry parameters in
(\ref{delta}) will lead to a non-zero variation of the action through
terms depending on the derivatives of $\e$. 

By just keeping track of the terms that depend on the (covariant)
derivatives of $\e$, it is straightforward to compute the
supersymmetry variation of the action on $M$ and the result is
(once again modulo total derivatives)
\be
\delta L_{SYM} = \left[(\nabla_{\mu}\bar{\e})\Ga^{NL}\Ga^{\mu}\Psi
           +\bar{\Psi}\Ga^{\mu}\Ga^{NL}(\nabla_{\mu}\e)\right]F_{NL}\non
\;\;.\label{m}
\ee
The $F_{NL}$-terms encapsulate the curvature terms $F_{\mu\nu}$ as
well as derivative terms of the scalars and scalar commutator terms.
Note that the supersymmetry parameters $\e$ are gauge
singlets so that the covariant derivative $\nabla_{\mu}\e$ includes
only the spin connection but not the gauge field. The gauge and
gravitational covariant derivative will be denoted by $D_{M}$.

\subsection{Killing Spinor Equations}

The most immediate non-trivial solutions to $\delta L_{SYM}=0$ 
(\ref{m}) are of course provided by parallel spinors, 
\be
\nabla_{\mu}\e=0\;\;.
\ee
The resulting supersymmetric theories and their Euclidean/topological
counterparts on Ricci-flat special holonomy manifolds are 
reasonably well understood
(see e.g.\ \cite{bt2,qmc2} and references therein)
and will not be considered further in this paper.

A natural generalization of a parallel spinors is a Killing spinor, i.e.\
a Dirac spinor $\eta$ in $n$ dimensions satisfying an equation 
of the form
\be
\nam\eta=\a\ga_{\mu}\eta
\label{kse}
\ee
where the $\ga_{\mu}$ are $n$-dimensional $\ga$-matrices and 
$\a$ is some real or imaginary constant.\footnote{Actually, while 
in the mathematics literature the name Killing spinor 
is usually reserved for spinors satisfying (\ref{kse}),
in the supergravity literature any equation of the form 
$\nabla_{\mu}\eta = M_{\mu}(x)\eta$
arising from setting to zero 
the gravitino variation in a bosonic background is 
called a Killing spinor equation. Here $M_{\mu}(x)$ is typically
made up from contractions of supergravity antisymmetric tensor
background fields with gamma matrices, hence the explicit $x$-dependence.
Here we have no such background fields, and thus we are left with
(\ref{kse}).} These equations have been
thoroughly investigated in the supergravity and mathematics literature,
at least in the case when $M$ is compact and Riemannian  - see e.g.\
\cite{bfgk,baer} and \cite{pr,vanN} and the references therein for
the mathematical and Kaluza-Klein supergravity aspects respectively.
For recent work on the pseudo-Riemannian case see \cite{bohle}.

To write this back in $d$-dimensional terms, it is not correct to just
consider an equation like $\nabla_{\mu}\e=\alpha\Ga_{\mu}\e$ as this
would for instance be incompatible with a chirality condition on $\e$.
Instead, we postulate the slightly more general Killing spinor equation
\be
\nabla_{\mu}\e = \alpha \Ga_{\mu}\Ga\e\;\;,
\label{gkse}
\ee
where $\e$ denotes the $d$-dimensional (chiral, Majorana, \ldots) spinor
and where $\Ga$ could be an arbitrary element of the Clifford algebra
generated by the $\Ga^{M}$. In fact we will be more specific than that and
consider the case in which $\Ga$ is a monomial constructed from the
`internal' gamma matrices $\Ga^{m}$, i.e.\ a completely anti-symmetrized
product of $0\leq p \leq d-n $ gamma matrices. When it is necessary to
indicate the degree $p$, we will write $\Ga^{[p]}$ instead of $\Ga$. Then
one in particular has $(\Ga)^{2}=\pm\II$. Generalizations of this are
certainly possible but will not be explored here.

This equation now preserves chirality when $p$ is odd, and so it can
also be used in the theories arising upon dimensional reduction of the
chiral $N=1$ theories.  Moreover, the freedom in the choice of $\Ga$
may allow one to find different supersymmetric theories for a given
field content (on manifolds satisfying either the same or different
integrability conditions of the Killing spinor equation). We will see
examples of this below.

Finally, this generalized Killing spinor equation, when written out in
$n$-dimensional terms, will always reduce to the standard Killing spinor
equation of the type (\ref{kse}) for (appropriate linear combinations of)
$n$-dimensional Dirac spinors\footnote{or perhaps to some simple variant
thereof when $n$ is even, 
\[ \nam\eta = i\a\ga_{\mu}\ga^{(n+1)}\eta \]
(here $\ga^{(n+1)}$ is the chirality operator). This equation can be
mapped to the standard equation (\ref{kse}) by passing to the unitarily
equivalent representation $\tilde{\ga}_{\mu}=i\ga_{\mu}\ga^{(n+1)}$.},
and therefore the standard existence criteria for ordinary Killing
spinors can be applied to (\ref{gkse}).

\subsection{Integrability Conditions}

The (first) integrability condition
arising from the Killing spinor equation (\ref{gkse}) is,
taking commutators and recalling (\ref{nabla}), 
\be
\trac{1}{4}\Ga_{ab}\Omega^{ab}_{\;\;\mu\nu}\e =
\a^{2}[\Ga_{\nu}\Ga,\Ga_{\mu}\Ga]\e\;\;,
\ee
where $\Omega^{ab}_{\;\;\mu\nu}$ denotes the curvature tensor of the
spin connection $\omega^{ab}_{\;\;\mu}$. 
Upon contraction with $\Ga^{\nu}$ this leads to 
\be
R_{\mu\nu}\Ga^{\nu}\e = -2 \a^{2}g_{\mu\nu}[(n-2)\Ga\Ga^{\nu}\Ga +
\Ga^{\nu}\Ga^{\la}\Ga\Ga_{\la}\Ga]\e\;\;.\label{ic}
\ee
For $\Ga=\Ga^{[p]}$ `internal' in the sense described before, so
that $\Ga^{[p]}$ commutes (anticommutes) with all the $\Ga^{\mu}$
if $p$ is even (odd), one finds 
\be
R_{\mu\nu}\Ga^{\nu}\e = -4\a^{2} (-1)^{p}(\Ga^{[p]})^{2} (n-1)
g_{\mu\nu}\Ga^{\nu}\e\;\;.
\label{ic3}
\ee
In the Riemannian case, an equation of the form $A_{\mu\nu}\Ga^{\nu}\e =0$
implies $A_{\mu\nu}=0$. This can be seen by multiplying by $A^{\mu}_{\;\la}
\Ga^{\la}$.
Thus in this case (\ref{ic}) implies that
\be
R_{\mu\nu} = -4\a^{2} (-1)^{p}(\Ga^{[p]})^{2} (n-1) g_{\mu\nu}\;\;.
\label{ic2}
\ee
and hence that $(M,g)$ is an Einstein manifold.
In particular, for $\Ga=\II$ or, equivalently,
for the ordinary Killing spinor equation (\ref{kse}), one obtains 
\be
R_{\mu\nu} = -4\a^{2} (n-1) g_{\mu\nu}\;\;.
\ee
Thus Killing spinors (\ref{kse}) for imaginary $\a$ (referred to as 
{\em real} Killing spinors in the mathematics literature) lead to
positive curvature, and spinors with real $\a$ (imaginary Killing 
spinors) lead to negative curvature.

This
unfortunate clash in terminology is due to the fact that typically in
the mathematics literature the conventions for Clifford algebras are
such that $\{\Ga_{\mu},\Ga_{\nu}\}=-2g_{\mu\nu}$, the opposite of the
convention used here. Perhaps a more invariant and informative
terminology would have 
been to call a Killing spinor positive or negative according to whether
the integrability condition leads to positive or negative curvature, and
we will adopt this terminology from now on. 

In general, the sign of the curvature depends on $\a$, $p$ and on
$(\Ga^{[p]})^{2}=\pm \II$. For the chiral $N=1$ theories and their
descendants, $p$ has to be odd in order for the Killing spinor equation
to be compatible with the chirality of $\e$. 

The integrability condition (\ref{ic2}) is not sufficient for the existence
of Killing spinors (not every Einstein manifold admits Killing spinors)
but fortunately an analysis of the higher integrability conditions can
be side-stepped by relating Killing spinors on $M$ to parallel spinors
on another Ricci flat manifold and hence establishing existence of 
Killing spinors directly  - see \cite{baer} for positive Killing spinors
and \cite{baum,bfgk} for negative Killing spinors.

In the pseudo-Riemannian case, (\ref{ic2}) is neither necessary nor
sufficient. An argument like the above 
only leads to the conclusion that for each value of $\mu$
the vector $V_{(\mu)}$ with components $V_{(\mu)}^{\;\nu}=A_{\mu}^{\;\nu}$
is null, with the additional constraint $g^{\mu\nu}A_{\mu\nu}=0$. In
the case of parallel spinors, the resulting {\em Ricci-null} Lorentzian
manifolds which are not Ricci flat were recently investigated in detail in
\cite{jffbmw}(see also \cite{bryant}). 
By the same token, one might suspect that there are
non-Einstein Lorentzian manifolds admitting Killing spinors. There are
indeed such examples for negative pseudo-Riemannian Killing spinors
whereas a pseudo-Riemannian manifold admitting a positive Killing spinor
is necessarily Einstein \cite{bohle}.  Nevertheless, in the following
we will simply assume that (\ref{ic2}) holds.  In this way we will
certainly miss some solutions (in the negative curvature case), but as
a first orientation this is good enough.

\subsection{The Supersymmetry Variation for Killing Spinors}

In order to plug (\ref{gkse}) into the formula (\ref{m}) for
$\delta L_{SYM}$, one first needs an expression
for $\nabla_{\mu}\bar{\e}$. By using the fact that
\be
(\Ga^{[p]})^{\dagger} = \eta_{p} A_{-}\Ga^{[p]} A_{-}^{-1}\;\;,
\ee
where 
\be
\eta_{p} = (-1)^{{p+1 \choose 2}}\;\;,
\label{etap}
\ee
one obtains
\be
\nabla_{\mu}\bar{\e}= -\eta_{p}\alpha^{*} \bar{\e} \Ga\Ga_{\mu}
\;\;.
\ee
Thus
\bea
\delta L_{SYM}&=&[-\eta_{p}\ab\eb \Ga \Ga_{\mu} \Ga^{NL}\Ga^{\mu}\Psi
+ \a\bar{\Psi} \Ga^{\mu}\Ga^{NL}\Ga_{\mu}\Ga\e]F_{NL}\non
&=& 2\mbox{Re}(\bar{\Psi}\Ga^{\mu}\Ga^{NL}\Ga_{\mu}\Ga\e))F_{NL}\;\;.
\eea
Splitting the gamma matrices $\{\Ga^{M}\} = \{\Ga^{\mu},\Ga^{m}\}$
and using the standard identities
\bea
\Ga_{\mu}\Ga^{\nu\lambda}\Ga^{\mu} &=& (n-4) \Ga^{\nu\lambda}\non
\Ga_{\mu}\Ga^{\nu m}\Ga^{\mu} &=& (n-2) \Ga^{\nu m}\non
\Ga_{\mu}\Ga^{lm}\Ga^{\mu} &=& n \Ga^{lm}\;\;,
\eea
one can evaluate this to find
\be
\begin{array}{cccll}
\delta L_{SYM}&=&
(n-4)&[-\eta_{p}\ab\eb \Ga \Ga^{\nu\lambda}\Psi + \a\bar{\Psi}
\Ga^{\nu\lambda}\Ga\e]&F_{\nu\lambda}\non
&+&2(n-2)&[-\eta_{p}\ab\eb \Ga \Ga^{\nu m}\Psi + \a\bar{\Psi} \Ga^{\nu m}\Ga\e]
&D_{\nu}\phi_{m} \non
&+&n&[-\eta_{p}\ab\eb \Ga \Ga^{lm}\Psi + \a\bar{\Psi}
\Ga^{lm}\Ga\e]&[\phi_{l},\phi_{m}]\;\;.\label{delta2}
\end{array}
\ee
Barring numerical coincidences, it is clear that this expression
can only vanish when the expression in brackets vanishes all by itself,
i.e.\ when 
\be
\mbox{Re}(\a \bar{\Psi}\Ga^{NL}\Ga \e)=0
\;\;\;\;\forall\;N,L\;\;.\label{m2}
\ee
This is only possible if $\a =0$ so that one is
dealing with ordinary parallel spinors (and hence Ricci flat geometries
in the Euclidean case and a few more possiblities for Lorentzian signature).

However, there is one numerical coincidence which occurs when $d=n=4$. In
that case only the first line of (\ref{delta2}) is present, but multiplied
by $n-4=0$.  Thus e.g.\ for any solution to the ordinary Killing spinor
equations (\ref{kse}) the $N=1$ theory in $d=3+1$ has a supersymmetry.
The relevant gamma matrix identity shows that this is due to the fact
that SYM theory is a theory of (non-Abelian) one-forms, and one might
want to speculate about an analogous result for (non-Abelian?) two-form
theories in $d=5+1$ \ldots

\section{Supersymmetric SYM Theories in Curved Space}

On the basis of these preliminaries we can now write down two families
of Dirac-Yang-Mills theories in curved space which are globally
supersymmetric courtesy of the existence of solutions to a suitable Killing
spinor equation. These theories generically differ from the simple SYM
action $L_{SYM}$ by mass terms for both the scalars and the fermions and by
a modified supersymmetry transformation rule for $\Psi$. In addition, one
class of these theories curiously has Chern-Simons-like cubic couplings
for the scalar fields.  Both of these families of theories turn out
({\em a priori} for no good reason) to be particularly simple in four
dimenions, $n=4$, and we will start with that particular case.

\subsection{Theories for $n=4$}

Let $L_{SYM}$ be the $(d,4)$ Lagrangian, that is the dimensional
reduction of the $d$-dimensional theory to $4=3+1$ dimensions, 
suitably covariantized, of course. Consider the Lagrangian 
\bea
L&=&L_{SYM} \mp 8\a^{2}\sum_{m=1}^{d-n}\phi_{m}^{2}\non
&=&-\trac{1}{2}F_{MN}F^{MN} + \bar{\Psi}\Ga^{M} D_{M}\Psi
\mp 8\a^{2}\sum_{m=1}^{d-n}\phi_{m}^{2}\;\;.
\eea
\label{l4}
This action is invariant under the supersymmetry transformations
(suppressing the Lie algebra labels on the fields)
\bea
\delta A_{M} &=& (\bar{\e}\Ga_{M}\Psi 
- \bar{\Psi}\Ga_{M}\e)\non
\delta \Psi  &=&  \Ga^{MN}\e F_{MN} -
4\a \sum_{m=1}^{d-n}\phi_{m}\Ga^{m}\Ga\e\;\;,
\label{susy4}
\eea
provided that $\e$ satisfies the Killing spinor equation
\be
\nam \e = \a\Ga_{\mu}\Ga\e
\ee
where $\Ga$ is any odd, internal matrix with $\Ga^{2}=\pm \II$. Here
$\a$ has to be real for the $d=4$ and $d=10$ Majorana(-Weyl) theories,
but can be either real or imaginary for the $d=4,6$ Weyl theories.

Indeed it is easy to see that due to the modification of the
$\Psi$-transformation the standard variation of $L_{SYM}$ given
in (\ref{delta2}) is exactly cancelled.  But now one picks up 
terms linear in the scalar fields $\phi_{m}$ from the Killing
spinor equation, namely when the derivative $D_{M}$ in the fermionic
kinetic term hits $\e$ in the second term of $\d \Psi$. 
This gives a term proportional to $\a^{2}$,
\be
\d L_{SYM} = \pm 16 \a^{2} [\bar{\e}\Ga^{m}\Psi -
\bar{\Psi}\Ga^{m}\e]\phi_{m}\;\;,
\ee
which is of course cancelled precisely by the variation of the 
mass term for the scalars.

Remarks:
\begin{enumerate}
\item We have just recovered the folklore statement that addition of
mass terms for the scalars is sufficient to render four-dimensional
SYM theories supersymmetric in a background admitting Killing spinors,
provided that also the supersymmetry transformation rules of the 
fermions are changed appropriately.
\item 
In particular, the mass term is precisely the conformally invariant mass term
arising in the conformally invariant wave operator
\be
\square -\frac{1}{4}\frac{n-2}{n-1}R\;\;,
\ee
where $R$ is the scalar curvature
\be
R=\pm 4\a^{2}n(n-1)\;\;.
\ee
\item Note the striking similarity of the supersymmetry transformations
with those of the special (i.e.\ superconformal)
supersymmetry transformations as given e.g.\ in 
\cite{mehta,chu,bilalchu}.
\item Similar linear terms in the transformations of the fermions 
also appear e.g.\ in the Wess-Zumino model in a curved background
\cite{bdfg} and are a rather generic feature of AdS supersymmetry
- for a recent review see \cite{bdewit}.
\item
Looking at the integrability conditions deduced before we learn
that in particular the counterpart of the
four-dimensional $N=2$ theory can be
supersymmetric on Einstein manifolds of either positive or negative
curvature admitting solutions of the Killing spinor equation, 
depending on whether $\a$ is chosen to be real or imaginary. 
\item 
Likewise, the 
$N=4$ theory can be supersymmetric in both cases, depending on
whether one chooses $\Ga=\Ga^{[1]}$ or $\Ga=\Ga^{[3]}$, with 
$\a$ real in both cases.  
\item
Even though the choice of $\Ga$ singles out one (or three) `internal'
directions, all the scalars have the same mass. This is a feature that
will not persist in $n\neq 4$. 
\item
There is no mass term for the fermions. Once again, this is a feature
peculiar to the $n=4$ theories. 
\item Finally, it may be possible to construct this theory as a rigid
limit of conformal supergravity in four dimensions.\footnote{I thank Ergin
Sezgin for this suggestion.}
\end{enumerate}

\subsection{Family A: Theories for $n\leq 5$ with $\Ga=\Ga^{[1]}$}

We will now consider the case where $\Ga$ is just a single internal
gamma matrix which we will call $\Ga^{1}$. In particular, 
$(\Ga)^{2}=+\II$. Now consider the following action
\be
L=L_{SYM} -4\a^{2}[(n-2)\sum_{m=1}^{d-n}\phi_{m}^{2} + (n-4)\phi_{1}^{2}]
-(n-4)\a\bar{\Psi}\Ga^{1}\Psi\;\;.
\label{l1}
\ee
As it stands this action makes sense for the $(d=4,n< 4)$, $(d=6,n<6)$,
and the fermionic mass term is hermitian provided that $\a$ is imaginary
(cf.\ the Appendix)
and this rules out the $(d=10, n\neq 4)$ theories.
We could also allow $(d=10,n=4)$ and $\a$ real, but in that case the
action reduces to the one discussed above. 

This action is invariant under the supersymmetry transformations
\bea
\delta A_{M} &=& (\bar{\e}\Ga_{M}\Psi 
- \bar{\Psi}\Ga_{M}\e)\non
\delta \Psi  &=&  \Ga^{MN}\e F_{MN} 
-4\a [\sum_{m=1}^{d-n}\phi_{m}\Ga^{m}\Ga^{1}\e + (n-4)\phi_{1}\e]
\label{susy1}
\eea
provided that $\e$ satisfies the Killing spinor equation
\be
\nam \e = \a\Ga_{\mu}\Ga^{1}\e
\ee

Remarks:
\begin{enumerate}
\item We now have mass terms both for the scalars and the fermions.
The masses depend only on the space-time dimension $n$, not on 
the parent dimension $d$. 
\item 
The mass of $\phi_{1}$ differs from that of the $\phi_{m\neq 1}$,
but neither is the conformally invariant value unless $n=2$ when
$d-n-1$ of the scalars are massless.
\item The integrability conditions tell us that these theories can only
exist on Einstein manifolds of negative curvature - in particular locally
AdS space-times. 
\item The $(6,5)$-theory on AdS${}_{5}$
has been constructed recently by Shuster \cite{sh} in terms of 
symplectic Majorana spinors. It can be checked that, when these
are reassembled into a Dirac spinor, his action and supersymmetries
agree precisely with those given above when one sets $d=6, n=5$.
\item The R-symmetry of the action has been reduced from $SO(d-n)$
(which is the manifest
R-symmetry group of the Poincar\'e supersymmetric theory)
to $SO(d-n-1)$.
\end{enumerate}

The simplest of these theories is the $(4,3)$ theory, i.e.\ the
$N=2$ theory in $n=3$. It differs from $L_{SYM}$ only by the fermionic
mass term, and the supersymmetry transformation rules are the standard ones,
i.e. we have
\bea
(d=4,n=3)&& L =L_{SYM} + \a \bar{\Psi}\Ga^{1}\Psi\non
         && \delta \Psi  =  \Ga^{MN}\e F_{MN} \;\;.
\eea
This theory is supersymmetric almost by inspection. For $n=3$, the 
two first lines of (\ref{delta2}) enter with opposite signs and 
the third line is absent. As $\Ga_{1}$ anticommutes with the 
$\Ga^{\mu}$ but commutes with the $\Ga^{\mu\nu}$, this is 
cancelled by the variation of the above fermionic  mass term.

Let us also write down explictly the $(6,3)$-theory. It is 
given by
\bea
(d=6,n=3)&& L =L_{SYM} -4\a^{2}(\phi_{2}^{2}+\phi_{3}^{2})
+ \a \bar{\Psi}\Ga^{1}\Psi\non
         && \delta \Psi  =  \Ga^{MN}\e F_{MN} -4\a(\phi_{2}\Ga^{2}+\phi_{3}
\Ga^{3}) \Ga^{1}\e\;\;.
\label{13}
\eea

\subsection{Family B: Theories for $n\leq 7$ with $\Ga=\Ga^{[3]}$}

If we want to use $\Ga=\Ga^{[3]}$ and still insist on this being
an element of the `internal' Clifford algebra, we obviously need
$n\leq d-3$. Let us choose 
\be
\Ga=\Ga^{123}=\frac{1}{3!}\e_{abc}\Ga^{abc}\;\;,
\ee
so that $(\Ga)^{2}=-\II$.  Consider the action 
\bea
L&=&L_{SYM} +4\a^{2}[(n-2)\sum_{m=1}^{d-n}\phi_{m}^{2} + (n-4)\sum_{a=1}^{3}
\phi_{a}^{2}]\non
&-&\frac{(n-4)\a}{3!}\e_{abc}\left[\bar{\Psi}\Ga^{abc}\Psi -
8\phi^{a}[\phi^{b},\phi^{c}]\right]\;\;.
\label{l3}
\eea
Hermiticity of the mass term requires $\a\in\RR$. 

This action is invariant under the supersymmetry transformations
\bea
\delta A_{M} &=& (\bar{\e}\Ga_{M}\Psi 
- \bar{\Psi}\Ga_{M}\e)\non
\delta \Psi  &=&  \Ga^{MN}\e F_{MN} 
-4\a [\sum_{m=1}^{d-n}\phi_{m}\Ga^{m}+ 
(n-4)\sum_{a=1}^{3}\phi_{a}\Ga^{a}] \Ga^{123}\e 
\label{susy3}
\eea
provided that $\e$ satisfies the Killing spinor equation
\be
\nam \e = \a\Ga_{\mu}\Ga^{123}\e
\ee

Remarks:
\begin{enumerate}
\item 
The most striking property of this action is perhaps the appearance of 
the cubic term for the scalar fields. It looks like the dimensional
reduction of a standard Chern-Simons term living in the three internal
directions singled out by $\Ga^{123}$. 
\item It is certainly suggestive of a supergravity origin of this term,
but it would  be desirable to find a pure gauge theory explanation 
for it as well.
\item Such terms can appear in the completely T-duality invariant D-brane
world-volume actions discussed by Myers in \cite{rcm}, where they arise
due to the coupling to non-trivial background antisymmetric tensor fields.
\item Some such term also appears in the off-shell 
{\em rheonomic} formulation 
of $N=1$ $d=10$ SYM in flat space - see \cite[(II.9.41)]{cdf}. The
relation to the appeareance of such a term 
in the on-shell space-time action  here is not clear (to the author)
but may be worth understanding. 
\item The integrability conditions once again lead to negative curvature
because even though $\a$ is now real, one also has $\Ga^{2}=-\II$. 
\end{enumerate}

Apart from the $(d=10,n=4)$ theory already discussed above, for which
there are neither fermionic mass terms nor Chern-Simons like couplings,
the simplest theory is once again the three-dimensional $(6,3)$-theory
with Lagrangian and supersymmetry transformation
\bea
(d=6,n=3)&& L =L_{SYM} + \a (\bar{\Psi}\Ga^{123}\Psi -
8\phi^{1}[\phi^{2},\phi^{3}])\non
         && \delta \Psi  =  \Ga^{MN}\e F_{MN} \;\;.
\label{susy33}
\eea
It is straightforward to check directly in this case that the 
action is supersymmetric: upon performing the supersymmetry variation,
the terms involving $F_{\mu\nu}$ and $D_{\mu}\phi_{n}$ arising 
from the variation of $L_{SYM}$ and the fermionic mass term cancel
whereas those involving the commutator $[\phi_{m},\phi_{n}]$ add up.
The latter are then precisely cancelled by the variation of the cubic
scalar term.
 
Note that we now have two obviously inequivalent curved space versions
of the $(6,3)$-theory, i.e.\ of what in standard parlance be called the
three-dimensional $N=4$ SYM theory ($N=4$ because in $2+1$ dimensions
spinors are two-component real: $so(2,1)\sim sl(2,\RR)$), one of them
with a standard mass term for two of the three scalars (\ref{13}), 
the other one instead with a Chern-Simons like term (\ref{susy33}). 
Is there any interesting
(duality?) relationship between these theories?  

\subsection{Euclidean Supersymmetric SYM Theories in Curved Space}

Euclidean (or better perhaps: Riemannian) versions of the theories
described above may be of interest for a variety of reasons, e.g.\
for D-brane instantons, within the Euclidean approach to the AdS/CFT
correspondence, and in connection with Hull's E-branes \cite{hulltimes}
and an eye towards cohomological versions of these theories.

As explained in \cite{bt2} (see also \cite{qmc2,bvn}), a convenient way
to obtain manifestly hermitian Euclidean SYM theories is by time-like
dimensional reduction of any one of the standard Minkowskian SYM
theories to a Lagrangian $L_{ESYM}$. 

This construction naturally explains the features one has in the past
come to expect of Euclidean supersymmetric theories, such as non-compact
R-symmetry groups (namely the internal roation group which is now the
Lorentz group $SO(d-n-1,1)$) and kinetic terms with the `wrong' sign
(namely the time-compoent of the gauge field, now a scalar from the
point of view of the Euclidean space-time).

These theories then also automatically make sense on Riemannian manifolds
and retain some fraction of their supersymmetry when this manifold
admits parallel spinors. In this way one obtains cohomological theories
on special holonomy manifolds with many beautiful features, studied for
example from this point of view in \cite{bt2,qmc2}.

Now let us, in analogy with what we did before, discuss the extension
of these Euclidean SYM theories to supersymmetric theories on Riemannian
manifolds admitting Killing spinors. Let us start with the $n=4$ theories
of section 3.1. It is readily seen that the theory as it stands is
supersymmetric also for the Euclidean theory provided that the mass
term is chosen to be $\sim \eta^{mn}\phi_{m}\phi_{n}$, i.e.\ 
\be
L=L_{ESYM} \mp 8\a^{2}\eta^{mn}\phi_{m}\phi_{n}\;\;,
\ee
for any choice of (internal, odd) $\Ga$. In particular, $\Ga$ could be
chosen to be equal to (or include) $\Ga^{0}$. The interesting thing
about this is that according to (\ref{ic2}) this changes the sign of
the integrability condition. Whereas for $\Ga=\Ga^{1}$, say, the sign
of the curvature is the sign of $\a^{2}$, for $\Ga=\Ga^{0}$ it is minus
the sign of $\a^{2}$.

This may not be of great consequence in the
present example since, as we saw before, we could anyhow obtain 
both signs by either choosing $\a$ to be real or imaginary (for 
the $(6,4)$ theory) or by choosing $\Ga=\Ga^{[1]}$ or $\Ga=\Ga^{[3]}$
(for the $(10,4)$ theory) - the integrability condition only depends on 
the square of $\a\Ga$. 

Moreover, for $n=4$, but only for $n=4$, there is practically no
dependence of the action on $\Ga$ (apart from the sign of the mass term)
so that we do not get any essentially new theories in this way. But we
will see below that in the other theories the freedom to choose $\Ga$
to include or not to include $\Ga^{0}$ gives us an added flexibility
not present in the pseudo-Riemannian theories. 

More or less the same modifications as above are required for the Family
A theories of section 3.2. Provided that we define the mass terms as
above and reintroduce the dependence of the sign of the mass term on
$\Ga^{2}=\pm \II$, as above, we obtain a supersymmetric Lagrangian. Thus
essentially the only two different possibilities are
\bea
\Ga = \Ga^{1}&& 
L=L_{ESYM} -4\a^{2}[(n-2)\eta^{mn}\phi_{m}\phi_{n} + (n-4)\phi_{1}^{2}]
-(n-4)\a\bar{\Psi}\Ga^{1}\Psi\non
\Ga = \Ga^{0}&& 
L=L_{ESYM} +4\a^{2}[(n-2)\eta^{mn}\phi_{m}\phi_{n} - (n-4)\phi_{0}^{2}]
-(n-4)\a\bar{\Psi}\Ga^{0}\Psi\non
\eea
We know that $\a$ has to be imaginary for hermiticity of the fermionic
mass term (this condition is the same for $\bar{\Psi}\Ga^{1}\Psi$
and $\bar{\Psi}\Ga^{0}\Psi$), and previously this forced the manifold
to have negative curvature. However, now we actually gain something
by being able to choose $\Ga=\Ga^{0}$ or $\Ga=\Ga^{1}$ (of course,
in order to have this choice one needs $n \leq d-2$).  Namely, the
Euclidean theory now has a supersymmetric version for negative curvature
($\Ga=\Ga^{1}$) and another supersymmetric version for positive curvature,
when $\Ga=\Ga^{0}$.

Mutatis mutandis one can draw the same conclusions for the theories of
section 3.3. The mass terms require the same treatment as before, and the
only novelty is the Chern-Simons-like cubic coupling for the scalar field.
If one chooses $\Ga = \Ga^{123}$, no further explanation is required. 
On the other hand, if one chooses, say, $\Ga=\Ga^{012}$, then one obviously
has to take into account the minus sign implicit in using
$\phi^{a}=\eta^{ab}\phi_{b}$. Thus explicitly the Chern-Simons term reads
\be
\frac{1}{3!}\epsilon_{abc}\phi^{a}[\phi^{b},\phi^{c}]=
-\phi_{0}[\phi_{1},\phi_{2}]\;\;.
\ee
The only thing worth noting here is perhaps that, unlike an ordinary
Chern-Simons term, which contains a first order time derivative,
this algebraic term remains real in Euclidean signature. The payoff 
from using $\Ga^{012}$ is that this theory exists on manifolds of 
positive curvature (admitting solutions of the corresponding Killing 
spinor equation, of course). Thus we have essentially the following
two Lagrangians:
\bea
\Ga=\Ga^{123}&& L=L_{ESYM} 
+4\a^{2}[(n-2)\eta^{mn}\phi_{m}\phi_{n} + (n-4)\d^{ab}
\phi_{a}\phi_{b}]\non
&&-(n-4)\a\left[\bar{\Psi}\Ga_{123}\Psi -
8\phi_{1}[\phi_{2},\phi_{3}]\right]\non
\Ga=\Ga^{012}&& L=L_{ESYM} 
-4\a^{2}[(n-2)\eta^{mn}\phi_{m}\phi_{n} + (n-4)\eta^{ab}
\phi_{a}\phi_{b}]\non
&&+(n-4)\a\left[\bar{\Psi}\Ga_{012}\Psi -
8\phi_{0}[\phi_{1},\phi_{2}]\right]\;\;.
\eea
We see that whereas in the pseudo-Riemannian case we had the freedom
to choose either positive or negative curvature space-times only for
$n=4$, in the Riemannian case the theories have this property for all $n$,
subject to the restrictions $n \leq d-2$ for the A theories and 
$n \leq d-4$ for the B theories. In $d-1$ (respectively $d-3$) dimensions,
there is no choice, $\Ga$ is dictated by whether one makes a purely
spaceklike or a (space-)time-reduction. 

\section{Aspects of the Supersymmetry Algebra}

In order to gain some insight into the structure of the theories
introduced above, and to attempt to understand them from the (A)dS
superalgebra point of view, in the following we will now (partially)
calculate the supersymmetry algebras in these models. 

\subsection{The Superalgebra for Family A}

Using (\ref{susy4}), it is straightforward to calculate 
the commutator of two supersymmetry transformations $\d_{i}$, 
associated with Killing spinors
$\e_{1}, \e_{2}$ satisfying $\nam\e_{i}=\a\Ga_{\mu}\Ga_{1}\e_{i}$, acting
on the bosonic fields $A_{\mu}$ and $\phi_{m}$. One finds
\bea
\trac{1}{4}[\d_{1},\d_{2}] A_{\mu} &=& 
V^{N}F_{N\mu}
+(n-3)(\a+\a^{*})V_{\mu}\phi_{1} + (\a-\a^{*}) V_{\mu i1}\phi^{i}\non
\trac{1}{4}[\d_{1},\d_{2}]\phi_{1}  &=& 
V^{N}F_{N1}
+(n-3)(\a+\a^{*})V_{1}\phi_{1} - (\a+\a^{*}) V_{i}\phi^{i}\non
\trac{1}{4}[\d_{1},\d_{2}]\phi_{j}  &=& 
V^{N}F_{Nj}
+(n-3)(\a+\a^{*})V_{j}\phi_{1} + (\a+\a^{*}) V_{1}\phi_{j}
+(\a-\a^{*}) V_{ij1}\phi^{i}\non
\label{dd1}
\eea
Here we have introduced the notation
\bea
V_{M}&=&\bar\e_{1}\Ga_{M}\e_{2}-\bar\e_{2}\Ga_{M}\e_{1}\non
V_{MNP}&=&\bar\e_{1}\Ga_{MNP}\e_{2}-\bar\e_{2}\Ga_{MNP}\e_{1}\;\;.
\eea
Ordinarily, i.e.\ in Poincar\'e supersymmetry, 
one would just find the first term on the right hand side.
Acting on the scalar fields, this is just the Lie derivative
(diffeomorphism) with resepect to $V^{\mu}$  plus a field dependent
gauge transformation, 
\bea
V^{N}F_{Nm}&=&L_{V}\phi_{m} + \d_{V}\phi_{m}\non
\d_{V}\phi_{m} &=& [V^{N}A_{N},\phi_{m}]\;\;.
\eea
Here and in the following it should be understood that the $V$ in
the Lie derivative refers only to the space-time components $V^{\mu}$
whereas all components $V^{M}$ enter in $\d_{V}$.

The same is true for the gauge field {\em provided that 
$V_{m}$ is constant}, as is the case for parallel spinors.
In that case, one has 
\bea
\nam V_{m}=0 \Ra && V^{N}F_{N\mu}=L_{V}A_{\mu} +\d_{V}A_{\mu}\non
                 && \d_{V}A_{\mu}=-D_{\mu}(V^{N}A_{N})\;\;.
\eea
However, when the $V_{m}$ are not constant, then one has instaed
\be
\nam V_{m}\neq 0 \Ra V^{N}F_{N\mu}=L_{V}A_{\mu} +\d_{V}A_{\mu} +
(\nam V_{m})\phi^{m}\;\;.
\ee
In order to understand how the right hand side of the supersymmetry 
algebra, including also all the other new terms,
nevertheless manages to be an invariance of the Lagrangian
in this case, we need to know some properties of the objects $V_{M}$
and $V_{MNP}$. The following identities are easily verified:
\bea
\nam V_{1} &=& - (\a+\a^*)V_{\mu} \non
\nam V_{i} &=& (\a-\a^*)V_{1i\mu} \non
\nam V_{ij1} &=& -(\a+\a^*)V_{\mu ij} \non
\nam V_{\nu} &=& (\a+\a^*)g_{\mu\nu}V_{1} + (\a-\a^{*})V_{1\nu\mu}\;\;.
\label{vid}
\eea
In particular, therefore, $V_{\mu}$ is a Killing vector if $\a^{*}=-\a$,
and a conformal Killing vector (and a gradient vector) if $\a^{*}=\a$.
In the former case, $V_{1}$ and the antisymmetric matrices $V_{1ij}$ 
are constant, whereas the other space-time scalars $V_{i}$ are not
(and vice-versa for $\a$ real). Moreover, note that the above 
equations imply that for $\a$ real the function $V_{1}^{2} + V_{\mu}V^{\mu}$
is constant. 

Using these results, 
we learn that the commutator of supersymmetry transformations
on the gauge field can be written as
\be
\trac{1}{4}[\d_{1},\d_{2}] A_{\mu} = L_{V}A_{\mu}+\d_{V} A_{\mu} -
(n-4) (\nam V_{1})\phi_{1}\;\;.
\ee
But since $V_{1}$ is constant for imaginary $\a$ and real $\a$ is 
only allowed for $n=4$, we see that in all cases the last term 
actually disappears and the commutator takes the standard form
\be
\trac{1}{4}[\d_{1},\d_{2}] A_{\mu} = L_{V}A_{\mu}+\d_{V} A_{\mu}\;\;.
\ee
If $\a$ is imaginary, then the commutator on the scalars takes the
form
\bea
\a^*=-\a \Ra && 
\trac{1}{4}[\d_{1},\d_{2}] \phi_{1} = L_{V}\phi_{1} +\d_{V} \phi_{1}\non
&&\trac{1}{4}[\d_{1},\d_{2}] \phi_{j} = L_{V}\phi_{j} +\d_{V} \phi_{j}
+2\a V_{ij1}\phi^{i}\;\;.
\eea
We see that in addition to diffeomorphisms (along a Killing vector)
and gauge transformations, the algebra now also includes a rotation
of the scalar fields by the constant matrix $V_{ij1}$ - this is 
(a subgroup of) the R-symmetry algebra of the theory and, combined with
an appropriate transformation of the fermions, a separate invariance of
the Lagrangian. The appearance of the R-symmetry algebra in the 
commutator of supersymmetries is of course a well known feature of
AdS superalgebras \cite{nahm} (for a recent review of AdS supersymmetry
see \cite{bdewit}) which we have recovered here somewhat experimentally.
Note that this extra rotation only appears for $n \leq d-3$. In particular,
it is absent for $n=4$.

The case $\a^*=\a$ (and thus $n=4$) is a bit more complicated, but this
should not be too surprising as now $V_{\mu}$ is only a conformal Killing
vector,
\be
L_{V} g_{\mu\nu} = 4 \a V_{1} g_{\mu\nu}\;\;,
\ee
and additional scale transformations of the scalars and fermions 
are required to produce an invariance of the Lagrangian density
in that case. Recall that precisely when $n=4$ the scalar field
action is conformally invariant so that this is feasible in principle.

The transformation on the gauge field is, as we have
noted above, the standard one, which is fine since the Yang-Mills 
Lagrangian is conformally invariant precisely when $n=4$. The scalars
now transform as
\bea
\a^*=+\a \Ra && 
\trac{1}{4}[\d_{1},\d_{2}] \phi_{1} 
= (L_{V}+2\a V_{1})\phi_{1}+\d_{V} \phi_{1}
 + \Delta_{V}\phi_{1}\non
&&\trac{1}{4}[\d_{1},\d_{2}] \phi_{j} 
= (L_{V}+2\a V_{1})\phi_{j}+\d_{V} \phi_{j}
 + \Delta_{V}\phi_{j}\;\;.
\eea
Here the modified Lie derivative $L_{V} + 2\a V_{1}$ reflects the 
non-trivial conformal weight of the scalar fields, and
\bea
\Delta_{V}\phi_{1}&=& -2\a V^{i}\phi_{i}\non
\Delta_{V}\phi_{j}&=& 2\a V_{j}\phi_{1}
\eea
is a particular global (the $V_{i}$ are constant in this case) 
infinitesimal $SO(d-4)$ rotation of the $(d-4)$ scalar
fields. This is only non-trivial for $d=6$ and for $d=10$. 
In the former case 
we find an $SO(2)$ rotation parametrized by $2 \a V_{2}$, namely
\bea
\Delta_{V}\phi_{1}&=& -2\a V_{2}\phi_{2}\non
\Delta_{V}\phi_{2}&=& 2\a V_{2}\phi_{1}\;\;.
\eea
Note that in this case ($\a$ real) the bosonic generators of the algebra
are conformal Killing vector fields that are also gradient vector fields
(this is an extremely restrictive condition but solutions exist e.g.\ in 
de Sitter space). As a consequence, since the Lie bracket of two gradient
vector fields is always zero, 
and also commutators of the modified operators 
$L_{V}+2\a V_{1}$ can be seen to vanish,
the bosonic part of the algebra engendered
in this way is Abelian, a situation apparently
not covered by Nahm's classification
\cite{nahm}. 

\subsection{The Complete Superalgebra for $n=4$}

Of course, to complete this discussion we should also calculate the
commutator of two supersymmetry transformations on the fermions. At
this point, because now Fierz identities are required, the discussion
becomes somewhat dimension-dependent and we
will only do this for $n=4$ which in many respects is the most 
interesting case to consider anyway.

For the $(6,4)$-theory, the supersymmetry variation of the spinor $\Psi$ 
is
\bea
\d \Psi &=& \Ga^{MN}\e F_{MN} - 4 \a \phi_{m}\Ga^{m}\Ga^{1}\e\non
        &=& \Ga^{MN}\e F_{MN} - 4 \a (\phi_{1} + \phi_{2}\Ga^{21})\e\;\;.
\eea
It follows that 
\bea
\d_{1}\d_{2}\Psi &=& 
2 D_{M}(\bar\e_1 \Ga_{N}\Psi-\bar{\Psi}\Ga_{N}\e_1) \Ga^{MN}\e_2
\non
&& - 4\a  (\bar\e_1 \Ga_{m}\Psi-\bar{\Psi}\Ga_{m}\e_1) \Ga^{m}\Ga^{1}\e_2
\;\;.
\eea
By the usual Fierz identity for SYM theories, the terms involving 
$\e_{1}$ and $\e_{2}$ will drop out after taking commutators and
we drop them henceforth. {}From the other terms we find, 
using the Killing spinor equation 
\be
\nam\bar\e = \a^{*}\bar\e\Ga_{1}\Ga_{\mu}\;\;,
\ee
that
\be
\d_{1}\d_{2}\Psi = 
2\a^{*}\bar\e\Ga_{1}\Ga_{\mu}\Ga_{N}\Psi \Ga^{\mu N}\e_2
+2 \bar\e_1 \Ga_{N}D_{M}\Psi\Ga^{MN}\e_2
- 4\a \bar\e_1 \Ga_{m}\Psi \Ga^{m}\Ga^{1}\e_2
\;\;.
\ee
Taking commutators and using the Fierz rearrangement formula 
for Weyl spinors $\Psi_{k}$ of the same chirality in $d$ dimensions, 
\be
\bar{\Psi}_{1}M \Psi_{2}\; \bar{\Psi}_{3}N \Psi_{4} = 
- 2^{-d/2}\sum_{p=0}^{n/2}c_{p}
\bar{\Psi}_{1}\Ga^{[p]}\Psi_{4} \; \bar{\Psi}_{3}N \Ga_{[p]} M\Psi_{2}\;\;,
\ee
(here a sum over the antisymmetrized products of $p$ gamma matrices
is understood) with 
\bea
c_{p}&=& (-1)^{p \choose 2}\frac{2}{p!} \;\;\;\;p < n/2\non
c_{n/2}&=& (-1)^{n/2 \choose 2}\frac{1}{(n/2)!} \;\;,
\eea
one obtains
\bea
{}[\d_{1},\d_{2}]\Psi &=& -\frac{1}{8} 2 \sum_{p} c_{p} 
(\bar\e_{1}\Ga^{[p]}\e_2 -\bar\e_{2}\Ga^{[p]}\e_1)\times\non
&\times& 
[ \Ga^{MN}\Ga_{[p]} \Ga_{N}D_{M}\Psi
+\a^{*}\Ga^{\mu N}\Ga_{[p]} \Ga_{1}\Ga_{\mu}\Ga_{N}\Psi 
- 2\a \Ga^{m}\Ga^{1}\Ga_{[p]} \Ga_{m}\Psi] \non
\label{fierz}
\eea
Now evidently only $p=1$ and $p=3$ contribute to the sum (this 
follows e.g.\ from the discussion leading to (\ref{a10})), 
giving rise to terms involving the vectors $V^{M}$ and antisymmetric
tensors $V^{MNP}$ encountered before. Upon using the
equation of motion $\Ga^{M}D_{M}\Psi = 0$, the first term will just 
give the standard contribution proportional to
\be
V^{M}D_{M}\Psi = V^{\mu}\nam \Psi + [V^{N}A_{N},\Psi]\;\;.
\ee
This has almost the right structure to be of the form diffeomorphism
plus gauge transformation we encountered for the bosonic fields. 
However, the (covariant) derivative on the spinor alone is not
part of the invariance of the action, i.e.\ the fermioic kinetic
term is not invariant under
\be
\d \Psi = V^{\mu}\nam \Psi
\ee
even if $V$ is Killing. Rather, for (conformal) Killing
vectors the Lie derivatives on the bosonic fields have to be supplemented
by the Lie derivative of the spinor field defined by
\be
L_{V}\Psi = V^{\mu}\nam\Psi + \frac{1}{4}\nam V_{\nu} \Ga^{\mu\nu}\Psi
\;\;.
\label{lvp}
\ee
Let us note here that in the present case the second term only contributes
when $V$ is a Killing vector ($\a$ imaginary), because $V$ is not only 
a conformal Killing vector but also a gradient
vector when $\a$ is real. This additional contribution to the covariant
derivative arises from the $p=3$ contributions to the second and third
terms in (\ref{fierz}) in the form
\be
\nam V_{\nu}\Ga^{\mu\nu} = (\a^* -\a) V_{\mu\nu 1}\Ga^{\mu\nu}\;\;.
\ee
The other $p=1$ contributions give rise to new terms in the 
supersymmetry algebra. After an altogether not particularly 
fascinating calculation one finds
\bea
\trac{1}{4}[\d_{1},\d_{2}]\Psi &=& L_{V} \Psi + \d_{V}\Psi \non
&& +\frac{1}{2}(\a + 5 \a^*) V_{1}\Psi 
+ \frac{1}{2}( \a + \a^{*})V_{i}\Ga^{i1}\Psi \;\;,
\eea
where $i\neq 1$.
Now let us take a look at this for $\a$ real and imaginary respectively.
For $\a$ imaginary, the complete commutator algebra reads
\bea
\trac{1}{4}[\d_{1},\d_{2}] A_{\mu} &=& L_{V} A_{\mu} + \d_{V}A_{\mu}\non
\trac{1}{4}[\d_{1},\d_{2}] \phi_{m}&=& L_{V} \phi_{m}+ \d_{V}\phi_{m}\non
\trac{1}{4}[\d_{1},\d_{2}]\Psi &=& L_{V} \Psi + \d_{V}\Psi -2\a V_{1}\Psi\;\;.
\label{s14ca}
\eea
Thus the only term we find in addition to the Lie derivative along
a Killing vector and a gauge transformation is a constant ($V_{1}$ is
constant) phase rotation ($\a$ is imaginary) of the spinor. The latter is
of course an invariance of the Dirac action - in fact it is the diagonal
$U(1)$ subgroup of the $SU(2)$ R-symmetry of the six-dimensional Weyl
action. It is nevertheless interesting that this additional phase
transformation appears in the commutator algebra for non-zero curvature.
Its appearance in the $(6,5)$ theory has been noted in \cite{sh}.

For $\a$ real, as we had seen before, already the algebra on the bosonic
fields is more complicated. In this case we have
\bea
\trac{1}{4}[\d_{1},\d_{2}] A_{\mu} &=& L_{V} A_{\mu} + \d_{V}A_{\mu}\non
\trac{1}{4}[\d_{1},\d_{2}] \phi_{1} &=& (L_{V}+2\a V_{1})\phi_{1}+\d_{V} \phi_{1}
 + \Delta_{V}\phi_{1}\non
\trac{1}{4}[\d_{1},\d_{2}] \phi_{j} 
&=& (L_{V}+2\a V_{1})\phi_{j}+\d_{V} \phi_{j} + \Delta_{V}\phi_{j}\non
\trac{1}{4}[\d_{1},\d_{2}]\Psi &=& (L_{V} + 3\a V_{1})\Psi + \d_{V}\Psi
+ \a V_{i}\Ga^{i1}\Psi\;\;.
\eea
Once again we see the modified Lie derivative on the spinor field (the 
factor of 3 reflecting the familiar conformal weight 3/2 of a spinor 
field). We also see the constant R-symmetry transformation 
\be
\Delta_{V}\Psi = \a V_{i}\Ga^{i1}\Psi
\ee
accompanying the rotation $\Delta_{V}\phi_{m}$ of the scalar fields. 
It is now straightforward to check that this indeed constitutes an 
invariance of the action, as it should. 

\subsection{The Superalgebra for Family B}

We will now calculate the action of the commutator of two
supersymmetry transformations on the bosonic fields for 
the family of Lagrangians (\ref{l3}) with supersymmetry
transformation (\ref{susy3}). Instead of $V_{M}$ and $V_{MNP}$,
this algebra will now contain in addition to the vector $V_{M}$
the rank five anti-symmetric tensor 
\be
V_{MNPQR} = \bar\e_1 \Ga_{MNPQR}\e_2 -\bar\e_2 \Ga_{MNPQR}\e_1\;\;.
\ee
A straightforward calculation gives
\bea
\trac{1}{4}[\d_{1},\d_{2}] A_{\mu} &=& 
V^{N}F_{N\mu} -2\a V_{123i\mu}\phi^{i}\non
\trac{1}{4}[\d_{1},\d_{2}]\phi_{a}  &=& 
V^{N}F_{Na}
+2(n-3)\a\e_{abc}\phi^{b}V^{c}\non
\trac{1}{4}[\d_{1},\d_{2}]\phi_{i}  &=& 
V^{N}F_{Ni}+ 2\a V_{123ij}\phi^{j}\;\;.
\label{dd3}
\eea
To interpret this, we proceed as in the analysis of (\ref{dd1}).
First of all we note the following properties:
\bea
\nam V_{a} &=& 0 \non
\nam V_{i} &=& 2\a V_{123i\mu}\non
\nam V_{\nu} &=& -2\a V_{123\mu\nu}\non
\nam V_{123ij} &=& 0\non
\nam V_{123} &=& 0\;\;.
\label{namv}
\eea
This shows that $V_{\mu}$ is a Killing vector and that the coefficients 
of the scalar field rotations are constants. There is an $SO(3)$
rotation acting on the three scalar fields $\phi_{a}$ and an 
$SO(d-n-3)$ rotation on the remaining scalars $\phi_{i}$.
The last relation, which we will only need later, shows that
$V_{123}$ is a constant, an imaginary constant to be precise.

Moreover, the second relation allows us to write, as before,  
\bea
\trac{1}{4}[\d_{1},\d_{2}] A_{\mu} &=& V^{\nu}F_{\nu\mu} -
V^{m}D_{\mu}\phi_{m} - (\nam V_{i})\phi^{i}\non
&=& L_{V}A_{\mu} + \d_{V}A_{\mu}\;\;,
\eea
so that all in all we have
\bea
\trac{1}{4}[\d_{1},\d_{2}] A_{\mu} &=& L_{V}A_{\mu} + \d_{V}A_{\mu}\non
\trac{1}{4}[\d_{1},\d_{2}] \phi_{a}&=& L_{V}\phi_{a}+ \d_{V}\phi_{a} 
+\Delta_{V}\phi_{a}\non
\trac{1}{4}[\d_{1},\d_{2}] \phi_{i}&=& L_{V}\phi_{i}+ \d_{V}\phi_{i} 
+\Delta_{V}\phi_{i}\;\;,
\eea
where
\bea
\Delta_{V}\phi_{a}&=& 2(n-3)\a\e_{abc}\phi^{b}V^{c}\non
\Delta_{V}\phi_{i}&=& 2\a V_{123ij}\phi^{j}\;\;.
\label{3phica}
\eea
Let us consider two special cases of this. The first is the 
$(6,3)$ theory. In this case evidently the commutator algebra
is just the standard algebra, in agreement with the fact that
the supersymmetry transformations themselves are just the
standard ones - see (\ref{susy33}). However, we will see below
that in spite of this the commutator algebra acting on $\Psi$
is different.

The second is the $(10,4)$ theory, i.e.\ the curved space 
counterpart of $N=4$ SYM theory in four dimensions. In this
case we see that the supersymmetry algebra exhibits an
$SO(3)\times SO(3)$ R-symmetry. I.e.\ from the point of 
view of the Poincar\'e supersymmetric theory the presence of
curvature has broken the R-symmetry down from $SO(6)$ to 
$SO(4)\sim SO(3)\times SO(3)$. This is in perfect agreement
with what a look at the AdS superalgebras would lead one to
conclude. The relevant superalgebra is now not the superconformal
$USp(N=4|4)$ with its $SU(4)$ R-symmetry but the AdS superalgebra
\be
OSp(N=4|4)\supset O(3,2) \times SO(4)\;\;.
\ee
It is rather pleasing to note that in the present 
context this reduction of the R-symmetry group can 
be traced back directly to the fact that the relevant 
Killing spinor equation involves the object $\Ga_{123}$.
This itself came from the requirement of having a hermitan
fermionic mass term for spinors that started off as
ten-dimensional Majorana-Weyl spinors.

\subsection{The Complete Superalgebra for $n=3$}

We have seen above that in the $(6,3)$ theory the supersymmetry
transformations (\ref{susy33}) and the supersymmetry commutator algebra
on the bosonic fields (\ref{3phica}) are just the usual ones, and one
might suspect that this essentially forces the commutator algebra on
the fermionic fields to be the standard one as well. However, this is
not necessarily the case.

First of all we know that the standard derivative term in the algebra
has to be promoted to the spinorial Lie derivative (\ref{lvp})
along a Killing vector field.

Secondly, in calculating $[\d_{1},\d_{2}]\Psi$ one encounters derivatives
of the spinor parameters and in this way the fact that the $\e_{i}$
are Killing spinors rather than parallel spinors feeds itself into
the algebra.

Thirdly, in calculating this algebra one makes use of the $\Psi$-equations
of motion. A look at the action (\ref{susy33}) reveals that these are
\be
\Ga^{M}D_{M}\Psi + \a\Ga_{123}\Psi = 0 \;\;,
\ee
and therefore no longer describe a massless spinor. 

And indeed one finds a new term in the commutator algebra even in
this case, where such a term is not required by similar terms
in the bosonic algebra. Starting from
\be
\trac{1}{4}[\d_{1},\d_{2}]\Psi = -\frac{1}{16}\sum_{p} c_{p} V^{[p]}
[ \Ga^{MN}\Ga_{[p]} \Ga_{N}D_{M}\Psi
-\a\Ga^{\mu N}\Ga_{[p]} \Ga_{123}\Ga_{\mu}\Ga_{N}\Psi] \;\;,
\ee
one finds that the first term contributes 
\be
\mbox{1st term} = \a V^{L}D_{L}\Psi +\frac{3}{8}\a V^{N}\Ga_{N}\Ga_{123}\Psi
+ \frac{1}{96}\a V^{[3]}\Ga_{[3]}\Ga_{123}\Psi\;\;,
\ee
which does not look particularly encouraging. 
However, the second term gives rise to 
\bea
\mbox{2nd term} = && \frac{5}{8}\a V^{\mu}\Ga_{\mu}\Ga_{123}\Psi
-\frac{3}{8}\a V^{a}\Ga_{a}\Ga_{123}\Psi\non
&& - \frac{3}{96}\a V^{[3]}\Ga_{[3]}\Ga_{123}\Psi
- \frac{1}{96}\a V^{[3]}\epsilon^{abc}\Ga_{a}\Ga^{\mu}\Ga_{[3]}\Ga_{\mu}
\Ga_{bc}\Psi\;\;.
\eea
The `mixed' three-index terms, i.e.\ those involving $V^{\mu\nu a}$
and $V^{\mu ab}$ cancel, while the other two, those involving 
$V^{\mu\nu\la}$ and $V^{abc}$, add up and (using the chirality of $\Psi$)
give rise to a single term proportional to $V_{123}\Psi$. 
The net result is then 
\be
\trac{1}{4}[\d_{1},\d_{2}]\Psi = V^{L}D_{L}\Psi + \a
V^{\mu}\Ga_{\mu}\Ga_{123}\Psi - 2\a V_{123}\Psi \;\;.
\ee
The second term is the missing contribution for the spinorial Lie
derivative (\ref{lvp}) as can be seen by using (\ref{namv}) and calculating
\bea
\frac{1}{4}\nam V_{\nu} \Ga^{\mu\nu}\Psi 
&=& -\frac{1}{2}\a V_{123\mu\nu}\Ga^{\mu\nu}\Psi \non
&=& \a V^{\mu}\Ga_{\mu}\Ga_{123}\Psi\;\;,
\eea
where the second equality follows from the chirality of $\Psi$.
Thus finally we have
\be
\trac{1}{4}[\d_{1},\d_{2}]\Psi = L_{V}\Psi + \d_{V}\Psi
- 2\a V_{123}\Psi \;\;,
\ee
and only the last term requires some conmment. As we have seen in
(\ref{namv}), $V_{123}$ is constant and, in fact, $(V_{123})^{\dagger}
=-V_{123}$, so that $V_{123}$ is an imaginary constant. But then
the Lagrangian (\ref{susy33}) is obviously invariant under this
phase rotation of the fermions. Once again, as in
(\ref{s14ca}), we find that the Killing spinor supersymmetry algebra 
includes this phase rotation for $\a \neq 0$, i.e.\ for curved spaces.

\section{The Coulomb Branch: Some Sample Calculations}

\subsection{Family A: Absence of a Maximally Supersymmetric Coulomb Branch}

Recall that the standard Poincar\'e supersymmetric SYM theory has the
Lagrangian
(\ref{lsym2})
\bea
L_{SYM} &=& 
-\trac{1}{2}F_{\mu\nu}F^{\mu\nu} -
D_{\mu}\phi_{m} D^{\mu}\phi^{m} 
- \trac{1}{2}[\phi_{m},\phi_{n}] [\phi^{m},\phi^{n}]\non
&+& \bar{\Psi}\Ga^{\mu} D_{\mu}\Psi
+ \bar{\Psi}\Ga^{m} [\phi_{m},\Psi] 
\eea
and the fermionic supersymmetry transformation (\ref{delta})
\be
\delta \Psi  =  \Ga^{\mu\nu}\e F_{\mu\nu} + 2\Ga^{\mu m}\e
D_{\mu}\phi_{m} + \Ga^{mn}\e [\phi_{m},\phi_{n}]\;\;.
\ee
The quartic potential has flat directions for mutually commuting scalar
fields. Thus there is a family of vacua parametrized by the constant
expectation values of the scalar fields taking values in the Cartan 
subalgebra of the gauge group (modulo the action of the Weyl group).
The supersymmetry transformations of the fermions are identically zero
in such a background without any condition on $\e$, and thus these
configurations parametrize a family of maximally supersymmetric vacua
of the SYM theory, the Coulomb branch.

We will now look for analogues of these
solutions in the Killing SYM theories we have discussed above, and we will see 
that typically (because of the modified supersymmetry transformations
and scalar potentials) there are no configurations which have all of
the above properties, but that there are half-supersymmetric configurations
which reduce to the above in the limit of vanishing curvature.

We begin by exploring the presence of a maximally supersymmetric purely
scalar field configuration in the theories of section 3. 
We will first consider the Family A theories for $d=6$ and 
$d=10$. 
The supersymmetry variation of the fermions in a purely scalar background 
becomes
\be
\d\Psi = 2\Ga^{\mu n}\e \del_{\mu}\phi_{n} + \Ga^{mn}\e[\phi_{m},\phi_{n}]
-4\a [\sum_{m=1}^{d-n}\Ga^{m}\Ga^{1}\e\phi_{m} + 
(n-4)\e \phi_{1}] \;\;.
\ee
It is clear almost by inspection that, unless $n=3$ and without any
further constraints on $\e$ beyond the chirality constraint dictated
by $d$-dimensional supersymmetry, vanishing of $\d\Psi$ implies
vanishing of all the $\phi_{m}$ because of the terms in $\d\Psi$ 
linear in the $\phi_{m}$. 

Indeed, first of all vanishing of the terms proportional to $\Ga^{\mu m}$
requires $\del_{\mu}\phi_{m}=0$. The term linear in $\phi_{1}$, 
proportional to the identity matrix acting on $\e$ has to vanish sepreately,
so one has $\phi_{1}=0$. The coefficient of $\Ga^{k1}$, $k\neq 1$, is
proportional to $[\phi_{k},\phi_{1}]-2\a\phi_{k}=-2\a\phi_{k}$, and therefore
also all the other scalar fields have to vanish, $\phi_{k}=0$. 

An exception occurs for $n=3$, as $\phi_{1}$ does then not appear in the term
in brackets proportional to $\a$ and can therefore be chosen to be
constant but otherwise unconstrained. By gauge invariance, this 
constant can be chosen to lie in the Cartan subalgebra of the gauge group.

Thus for $n\neq 3$ there are no non-trivial 
maximally supersymmetric purely scalar configurations (switching on any
scalar vev breaks at least some fraction of the supersymmetry), while for
$n=3$ there is (for $G=SU(2)$) a one-dimensional Coulomb `twig'. 

If these gauge theories can be shown to
arise as worldvolume theories of branes,
this should have implications for the possibility (or lack thereof) to
move them apart, and thus also for the question of existence of marginal
bound states among these branes. 

\subsection{Family B: Existence of a Discrete Family of Maximally 
Supersymmetric Scalar Field Configurations}

For the Family B theories, with their supersymmetry variation
\be
\delta \Psi  =  \Ga^{MN}\e F_{MN} 
-4\a [\sum_{m=1}^{d-n}\phi_{m}\Ga^{m}+ 
(n-4)\sum_{a=1}^{3}\phi_{a}\Ga^{a}] \Ga^{123}\e 
\;\;,
\ee
the situation is somewhat different.

In particular, as we had seen in (\ref{susy33}), $\a$
disappears altogether from the supersymmetry transformation
rules for $(d=6,n=3)$. In that particular case, we therefore find
the `normal' Coulomb branch parametrized by the three constant
commuting scalars. These solutions are also the only maximally
supersymmetric critical points of the scalar cubic plus quartic potential.

For the reductions of the $d=10$ theories to $n\leq 7$ dimensions the
situation is the following. We once again set the gauge fields to zero.
Then imposing $\d\Psi=0$ forces the scalars to be constants.
The terms linear in the $\phi_{k}$, 
$k\neq 1,2,3$ are proportional to $\Ga^{k123}\e$ and have to vanish
seperately. Thus $\phi_{k}=0$. For the remaining scalar fields $\phi_{a}$,
by looking at the coefficients of $\Ga^{ab}\e$ we find the condition
\be
[\phi_{a},\phi_{b}]=2\a(n-3) \epsilon_{abc}\phi_{c}
\;\;.
\label{su2}
\ee
Up to an irrelevant scaling, this amounts to a 
homomorphism of the Lie algebra of $SU(2)$ into that of 
the gauge group $G$ and hence there are maximally supersymmetric
vacua for each conjugacy class of such homomorphisms.  

It can also be checked directly that this gives a critical point
of the potential (with $\phi_{m}=0$ for $m\neq 1,2,3$)
\be
V(\phi)=-\trac{1}{2}\Tr[\phi_{a},\phi_{b}]^{2} + 8\a^{2}(n-3)\Tr\phi_{a}^{2}
+\trac{4}{3}\a(n-4)\epsilon_{abc}\phi_{a}[\phi_{b},\phi_{c}]\;\;.
\ee

This is reminiscent of the analysis by Vafa and Witten \cite{vw}
of the vacua of the mass-perturbed $N=4$ SYM theory: in that case 
the cubic superpotential of the $N=4$ theory (in $N=1$ language)
is perturbed by quadratic mass terms, and the equation for the critical
points is equivalent to (\ref{su2}).\footnote{For a recent discussion
of these theories in the context of the AdS/CFT correspondence see
\cite{jpms}.} Here we find this solution even
in the presence of an additional quartic term in the potential.

We see that for these theories there are indeed maximally supersymmetric
vacua, but that their structure is rather different from that of the
standard Coulomb branch. Instead of a continous we have a discrete
family of vacua with unbroken supersymmetry, and this is reflected in
the absence of flat directions in the scalar potential for $n\neq 3$.

\subsection{Existence of a half-BPS Coulomb Branch for AdS Space-Times}

In order to study configurations preserving some fraction of the
supersymmetry, we need to know what kind of additional conditions
can be imposed on an $n$-dimensional Killing spinor. Clearly a 
chirality condition (which is the natural condition for constant
or parallel spinors) is incompatible with the Killing spinor equation
\be
\nam\eta = \a \ga_{\mu}\eta\;\;.
\ee
Fortunately, very compact and explicit 
expressions are known \cite{lpt} for Killing spinors on AdS
space-times, and these results will enable us to find half-supersymmetric
scalar field configurations. 

We begin by quickly reviewing the results obtained in \cite{lpt}.
The $AdS_{n}$ metric takes a particularly simple form in horospheric
(or the closely related Poincar\'e) coordinates, in which one 
has 
\be
ds^{2} = dr^{2} + \ex{\frac{2r}{\ell}}\eta_{ij}dx^{i}dx^{j}\;\;.
\ee
The scalar curvature of this metric is 
\be 
R = - \frac{1}{\ell^{2}}n(n-1)\;\;,
\ee
which identifies $\ell$ as the curvature radius of the space-time, related 
to our constant $\a$ by $|\a|= 1/2\ell$. 
The spinorial covariant derivative
in these coordinates is
\bea
\nabla_{r}\eta &=& \del_{r}\eta\non
\nabla_{k}\eta &=& \del_{k}\eta +\frac{1}{2\ell}\ga_{k}\ga_{r}
\eta\;\;.
\eea
Hence the Killing spinor equation 
\be
\nam \eta = \frac{1}{2\ell}\ga_{\mu}\eta
\ee
can be written as the pair of equations
\bea
\del_{r}\eta &=& \frac{1}{2\ell}\ga_{r}\eta\non
\del_{k}\eta &=& \frac{1}{2\ell}\ga_{k}(1-\ga_{r})\eta\;\;.
\eea
Clearly, if $\ga_{r}\eta=\eta$, the solutions are
\be
\eta^{+}=\ex{\frac{r}{2\ell}}\eta^{+}_{0}\;\;,
\ee
where $\eta^{+}_{0}$ is an arbitrary constant spinor satisfying
\be
\ga_{r}\eta^{+}_{0}= \eta^{+}_{0}\;\;.
\ee
These are the Killing spinors we will consider in the following. 
The general solution is
\be
\eta= \ex{\frac{r}{2\ell}\ga_{r}}
(1+ \frac{1}{2\ell}x^{k}\ga_{\ul{k}}(1-\ga_{r}))
\eta_{0}\;\;,
\ee
where $\eta_{0}$ is now an arbitrary constant spinor and 
$\ga_{\ul{k}}$ refers to an orthonormal basis. This shows 
that AdS has the maximal number of linearly independent Killing spinors,
i.e.\ is maximally supersymmetric in the supergravity sense. 

Armed with these solutions to the Killing spinor equations, we
can now reconsider the issue of supersymmetric purely scalar field
configurations. For concreteness we consider the Family A $(6,n)$
theories for $n=4$ and $n=5$.

For $AdS_{5}$ we choose gamma-matrices $\ga_{k}, k=0,1,2,3$
satisfying 
\be
\{\ga_{k},\ga_{l}\} = \ex{\frac{2r}{\ell}}\eta_{kl}
\ee
and $\ga_{r}=\ga^{(5)}$. A convenient basis for the $d=6$
Clifford algebra is then 
\bea
\Ga_{k} &=& \sigma_{1}\otimes \ga_{k}\;\;\;\; k = 0,\ldots,n-2=3\non
\Ga_{r} &=& \sigma_{1}\otimes \ga^{(5)}\non
\Ga_{5} &=& \sigma_{2} \otimes \II
\eea
where we have now, for sanity's sake, called the internal gamma matrix
appearing in the Killing spinor equation
\be
\nam\e = \a\Ga_{\mu}\Ga_{5}\e\;\;,
\ee
$\Ga_{5}$ instead of $\Ga_{1}$. 
For $n=4$ we will choose a dimensional reduction along the $x^{3}$-direction
so that now $\{\ga_{\mu}\}=\{\ga_{k},\ga^{(5)}\}$ with $k=0,1,2$.

For $\e$ a six-dimensional Weyl spinor, $\e^{T}=(\eta^{T}, 0)$
the Killing spinor equation reduces to
\be
\nam\eta=i\a\ga_{\mu}\eta
\ee
so we have the identification 
\be
i\a = \frac{1}{2\ell}\;\;.
\ee
Therefore the AdS Killing spinor equation becomes 
\bea
\nabla_{k}\eta&=&\frac{1}{2\ell}\ga_{k}\eta\non
\nabla_{r}\eta&=&\frac{1}{2\ell}\ga^{(5)}\eta\;\;,
\eea
so that indeed $\ga_{r}=\ga^{(5)}$ and the condition 
$\ga_{r}\eta=\eta$ translates into a 
standard chirality condition in the four-dimensional sense.

We begin with the $n=5$ theory, denote the single scalar field
simply by $\phi$, and consider the fermionic variation (once again,
we set the gauge fields to zero)
\be
\d\Psi = 2\Ga^{k5}\e\del_{k}\phi + 
2\Ga^{r5}\e \del_{r}\phi -8\a\phi\e\;\;.
\ee
Translating this into five-dimensional gamma matrices acting on
$\eta$, one finds
\be
\d\Psi = 0 \LRa 2i\ga^{k}\eta\del_{k}\phi 
+ 2i\ga_{r}\eta\del_{r}\phi + \frac{4i}{\ell}\eta\phi
=0 \;\;.
\ee
Now we find that for Killing spinors satisfying 
$\ga_{r}\eta=\eta$, the supersymmetry condition becomes
$\del_{k}\phi=0$ and
\be
\del_{r}\phi=-\frac{2}{\ell}\phi\;\;,
\ee
or
\be
\phi = \ex{-\frac{2r}{\ell}}\phi^{0}\;\;,
\ee
where $\phi^{0}$ is an arbitrary constant anti-hermitian matrix in the Lie 
algebra of the gauge group. 

Let us note the following properties of this configuration:
\begin{enumerate}
\item By construction, this configuration leaves half of the supersymmetries
(namely those associated with Killing spinors satisfying 
$\ga_{r}\eta=\eta$) unbroken.
\item It is also a solution to the equations of motion. The equation of
motion is (with the mass term expressed in terms of $\ell$)
\be
\square \phi = -\frac{4}{\ell^{2}}\phi\;\;.
\ee
On functions depending only on $r$, this reduces to 
\be
(\del_{r}^{2}+ \frac{4}{\ell}\del_{r})\phi= 
-\frac{4}{\ell^{2}}\phi\;\;,
\ee
which is satisfied by $\phi\sim \exp(-2r/\ell)$. 
\item In the flat space limit $\ell\ra\infty$, $\phi$ just reduces
to a constant. In that limit there is a supersymmetry enhancement
and $\phi^{0}$ parametrizes the maximally supersymmetric Coulomb branch of 
the five-dimensional $N=2$ theory. 
\end{enumerate}

For $n=4$ the situation is quite similar. We now have two
scalar fields which, with the above conventions, would 
most naturally be called $\phi_{3}$ (say) and $\phi_{5}$.
But I will just call them $\phi_{1,2}$.
Vanishing of the supersymmetry transformation in this
case (for the $\ga_{r}=+1$ Killing spinors)
forces these fields to be $x^{k}$-independent 
and to commute, and the $r$-dependence is determined by
\be
\del_{r}\phi_{1,2}=-\frac{1}{\ell}\phi_{1,2}\;\;,
\ee
leading to
\be
\phi_{1,2} = \ex{-\frac{r}{\ell}}\phi^{0}_{1,2}\;\;.
\ee
These are once again half-supersymmetric solutions to the equations
of motion, which in this case read
\be
(\del_{r}^{2}+ \frac{3}{\ell}\del_{r})\phi_{1,2}= 
-\frac{2}{\ell^{2}}\phi_{1,2}\;\;,
\ee
and tend to the standard Coulomb branch of $N=2$ $n=4$ SYM as
$\ell\ra\infty$. Once again in that limit one finds a supersymmetry
enhancement.  

\section{Open Issues: Interpretation and Applications}

Above we have constructed two families of curved space counterparts
of the standard Poincar\'e supersymmetric SYM theories which are
globally supersymmetric on manifolds admitting Killing spinors, and
we also began a preliminary investigation of their properties. But
clearly a large number of issues still remain to be understood.  

\begin{enumerate}
\item 
Foremost among them is perhaps the relevance of these theories to
the dynamics of D-branes. For this one might also want to consider
spacetimes of the form $M=\Sigma\times\RR$ where $\Sigma$ admits
Killing spinors. The analysis closely resembles the one for Euclidean
theories on $\Sigma$ described in section 3.4.

If these theories play a role in that context, what are the consequences 
of the unusual properties of the Coulomb branch we have found in 
section 5? Where would one expect the mass or cubic potential terms
to show up in applications? What about BPS configurations with non-trivial
gauge fields (monopoles) in these theories? What is the relation to the BPS 
configurations in AdS space studied e.g.\ in \cite{yi,grst}? 
What is the relation to the AdS calibrations of \cite{jggp,gpt}?
Are there interesting cohomological versions of these theories?

\item One might also want a better understanding of the superalgebras
underlying these theories, depending on the number of available
Killing spinors.  What about the $(d=10,n=8,9)$ theories? How is the 
problem to construct such theories related to the absence of 
conventional AdS superalgebras beyond $n=7$? What about central charges
and the addition of matter fields?

\item It would also be desirable to have a more conceptual understanding
of the existence of these two classes of theories. For the Family A
theories a possible approach may be the following. There is a one-to-one
correspondence between (Riemannian, positive) Killing spinors on $M$
and parallel spinors on the so-called cone $CM$ over $M$ \cite{baer}
(see e.g.\ \cite{qmc3} for a survey of these matters in the AdS/CFT
context), with similar results for other signatures and signs. Thus the
parallel spinors on $CM$ appear to play a dual role.  On the one hand,
they assure the supersymmetry of SYM theory on $CM$. On the other hand,
they are invoked to establish the existence of Killing spinors on $M$
and hence supersymmetry of SYM theory on $M$. It is therefore natural to
wonder if these two appearances of parallel spinors are related and if,
indeed, a straightforward dimensional reduction of the supersymmetric
theory on $CM$ might not have been a less roundabout way of arriving at
the theory on $M$.

The problem with a naive dimensional reduction of a theory on $CM$ to
one on $M$ is that there is no isometry in the cone direction but only
a homothety.  This  suggests that perhaps one way to reduce a theory on
$CM$ to a theory on $M$ is to perform a Scherk-Schwarz like reduction
or gauging along the radial direction.  The structure of the Family A
theories is certainly suggestive: one `internal' gamma-matrix $\Ga_{1}$
is singled out, which should be identified with $\Ga_{r}$, and the mass
terms could arise from a Scherk-Schwarz like reduction. However, so far
I have been unable to derive these theories in this way.

\item For the theories in Family B, an altogether different idea appears
to be required to account for the Chern-Simons-like terms.  
The appearance of such a term
in the $n$-dimensional action suggests an $(n+3)$-dimensional origin 
with a true CS term living in those extra three dimensions. Thus one
should have a coupling 
\[\int F^{(n)}(AdA+\ldots)\]
where $F^{(n)}$ is proportional to the volume form on $M$. Thinking
of this as a RR field strength, one recognizes the Wess-Zumino 
coupling of a $D(n+2)$-brane world volume to a $D(n-2)$-brane via the
instanton action $\Tr F\wedge F$. E.g.\ for $n=5$ and 
$AdS_{5}$ one has a $D3-D7$ brane system. And indeed in the near-horizon
limit of such a system one obtains $AdS_{5}\times X_{5}$, where 
$X_{5}=S^{5}/\ZZ_{2}$ has a fixed $S^{3}$ over which the $D7$-branes 
are wrapped \cite{sgd,apty} and the $F^{(5)}$ is proportional to
the volume element on $AdS_{5}$ (plus its Hodge dual). 
Thus the $D7$-$O7$ couplings of the form
\[\int C^{(4)}\wedge\Tr F\wedge F\]
could be responsible for the Chern-Simons like terms in the five-dimensional
gauge theory obtained by reduction of the worldvolume theory of the 
$D7$-branes to $AdS_{5}$.

Of course, even if one can trace the Chern-Simons terms back to these
configurations (and hence the corresponding supergravity theory), 
one still needs to understand why they are required by
supersymmetry for a gauge theory on $AdS_{n}$ (or some other space-time
admitting Killing spinors). However, perhaps the above considerations
may at least provide a first step to such an understanding.

Alternatively, the existence of such terms in the action could be deduced
from considerations as in \cite{rcm}, where D-brane actions in non-trivial
antisymmetric tensor field backgrounds (and hence also non-trivial curvature
by the Einstein equations) are studied. 
\end{enumerate}

\appendix

\section{Some Useful Identities for Fermion Bilinears}

To understand the hermiticity properties of 
fermionic mass terms, which play an important role 
in the discussion of section 3, 
and in order to facilitate other
manipulations, it is useful to know some identities for spinor bilinears 
involving gamma-matrices. First of all, let us introduce the unitary
matrices $A_{\pm},B_{\pm},C_{\pm}$ by
\bea
\Ga_{M}^{\dagger}&=&\pm A_{\pm}\Ga_{M}A_{\pm}^{-1}\non
\Ga_{M}^{*}&=&\pm B_{\pm}\Ga_{M}B_{\pm}^{-1}\non
\Ga_{M}^{T}&=&\pm C_{\pm}\Ga_{M}C_{\pm}^{-1}\;\;.
\eea
We can always choose $A_{-}=\Ga_{0}=-A_{-}^{\dagger}$, and for 
$d$ even for $A,B$ and $C$ the $\pm$ matrices are related by 
multiplication by $\Ga^{(d+1)}$. For a general analysis see 
e.g.\ \cite{kt}.

Majorana spinors are characterized by the condition
\be
\Psi^{*}=B_{\pm}\Psi\;\;,
\ee
which is consistent provided that
\be
B_{\pm}^{*}B_{\pm}=\II\;\;.
\ee
Then for a Majorana spinor one has
\be
\bar{\Psi}=\Psi^{\dagger}A_{-}=\Psi^{T}B_{\pm}^{T}A_{-}\;\;.
\ee
But one can easily check that, given the properties of $A$ and $B$,
one has
\be
B_{\pm}^{T}A_{-}\Ga_{M}(B_{\pm}^{T}A_{-})^{-1}=\mp\Ga_{M}^{T}\;\;,
\ee
and thus one can identify
\be
C_{\mp}=B_{\pm}^{T}A_{-}\;\;.
\ee
Hence the Majorana condition can also be written as
\be
\bar{\Psi}=\Psi^{T}C_{\mp}\;\;,
\ee
which is perhaps more familiar. For the Majorana(-Weyl) theories in
$d=3+1$ and $d=9+1$, we will usually choose $B=B_{+}$ to obtain
\be
B=B_{+}\Ra \bar{\Psi}=\Psi^{T}C_{-}\;\;.
\ee
In a Majorana basis of real gamma-matrices, 
one can always choose $B_{+}=\II$ and $A_{-}=C_{-}$,
since $\Ga_{M}^{\dagger}=\Ga_{M}^{T}$ and hence Majorana spinors
are real in such a basis.

Now let us look quite generally at a spinor bilinear
\be
\bar{\Psi}\Ga^{[p]}\Phi\;\;.
\ee
If $\Psi$ and $\Phi$ are chiral spinors, then it is easy to see that
this bilinear is zero if $p$ is even and $\Psi$ and $\Phi$ have the 
same chirality (and likewise is zero if $p$ is odd and $\Psi$ and $\Phi$
have opposite chiralities). To see this one can calculate, using
$\Ga^{(d+1)\dagger}=\Ga^{(d+1)}{}^{-1}=\Ga^{(d+1)}$, 
\be
\overline{\Ga^{(d+1)}\Psi}\Ga^{[p]}\Phi=(-1)^{p+1}\bar{\Psi}\Ga^{[p]}
\Ga^{(d+1)}\Phi;\;,
\label{a10}
\ee
from which the claim follows. Now let us check under which conditions 
the corresponding mass term is hermitian. To that end we calculate, noting an
extra minus sign due to working with anticommuting spinors,
\be
(\bar{\Psi}\Ga^{[p]}\Phi)^{\dagger}=\eta_{p}\bar{\Phi}\Ga^{[p]}\Psi
\ee
($\eta_{p}$ was defined in (\ref{etap}))
so that $\bar{\Psi}\Ga^{[p]}\Psi$ is hermitian for $\eta_{p}=+1$,
i.e.\ $p=0,3,4,7,8\ldots...$ while for $\eta_{p}=-1$, one has to multiply 
this term by $i$ to obtain a hermitian mass term. 

For $\Psi$ and $\Phi$ Majorana, one has, using also $C^{T}=-C$ (in a 
Majorana basis)
\be
\bar{\Psi}\Ga^{[p]}\Phi=(\bar{\Psi}\Ga^{[p]}\Phi)^{T}=
\eta_{p}\bar{\Phi}\Ga^{[p]}\Psi
\ee
consistent with the fact that in a Majorana basis transposition and 
hermitian conjugation are the same operation. Thus the potential
mass term $\bar{\Psi}\Ga^{[p]}\Psi$
is zero unless $\eta_{p}=+1$ (and in this case we are not permitted to 
render the mass term hermitian for $\eta_{p}=-1$ by multiplying it
by $i$). 

Summarizing the above discussion, we see that for the $d=2+1$ Majorana
theory, the only posibility is $p=3$, equivalent to $p=0$ because
$\Ga_{012}$ is a multiple of the identity in that case.  Likewise, for
the $d=3+1$ Majorana theory, the only possibilities are $p=0,3,4$. For the
chiral version of that theory, we have $p=1$ or $p=3$ (with imaginary and
real coefficients respectively, related to the fact that $\Ga^{(5)}$ has a
factor of $i$).  For the chiral theory in $d=5+1$, one necessarily has $p$
odd, and therefore either $p=1$ (equivalent to $p=5$) with a factor of
$i$, or $p=3$ with a real coefficient. We will find supersymmetric gauge
theories for either choice of mass term.  Finally, the only possibility
for the Majorana-Weyl theory in $d=9+1$ is $p=3$.

\rnc{\Large}{\normalsize}

\end{document}